\documentclass[aps,pre,10pt,twocolumn,showkeys,
               superscriptaddress]{revtex4-2}

\usepackage[normalem]{ulem}
\usepackage{amsmath,amsfonts}
\usepackage{color,physics,graphicx,bm}
\usepackage{hyperref}
\hypersetup{colorlinks,citecolor=red,filecolor=black,linkcolor=blue,urlcolor=black}

\newcommand{\ii}{\mathrm{i}}

\begin{document}
\title{Dephasing-induced relaxation in tight-binding chains 
with linear and nonlinear defects}
\author{Debraj Das}
\email{ddas@sissa.it}
\affiliation{SISSA -- International School for Advanced Studies, Via Bonomea 265, 34136 Trieste, Italy}
\affiliation{Istituto dei Sistemi Complessi, Consiglio Nazionale delle Ricerche, via Madonna del Piano 10, 50019 Sesto Fiorentino, Italy}

\author{Andrea Gambassi}
\email{gambassi@sissa.it}
\affiliation{SISSA -- International School for Advanced Studies, Via Bonomea 265, 34136 Trieste, Italy}
\affiliation{Istituto Nazionale di Fisica Nucleare, Sezione di Trieste, via Valerio 2, 34127 Trieste, Italy}

\author{Stefano Iubini}
\email{stefano.iubini@cnr.it}
\affiliation{Istituto dei Sistemi Complessi, Consiglio Nazionale delle Ricerche, via Madonna del Piano 10, 50019 Sesto Fiorentino, Italy}
\affiliation{Istituto Nazionale di Fisica Nucleare, Sezione di Firenze, via G. Sansone 1, 50019 Sesto Fiorentino, Italy}

\author{Stefano Lepri}
\email{stefano.lepri@cnr.it}
\affiliation{Istituto dei Sistemi Complessi, Consiglio Nazionale delle Ricerche, via Madonna del Piano 10, 50019 Sesto Fiorentino, Italy}
\affiliation{Istituto Nazionale di Fisica Nucleare, Sezione di Firenze, via G. Sansone 1, 50019 Sesto Fiorentino, Italy}

\date{\today}

\begin{abstract}
We investigate thermalization in a tight-binding chain with an on-site defect subject to local dephasing noise implemented as random phase kicks. For a single linear defect of strength $\epsilon$, we obtain an exact analytical description of the system spectrum and formulate the dephasing-induced dynamics in the eigenstate basis. We derive an approximate kinetic equation for mode populations that describes a continuous-time random walk in action space. The walk transition rates are defined by the overlap matrix encoding the spatial structure of eigenstates that can be computed exactly. Analyzing the spectral properties of the equation, we show that defect-induced localized modes act as bottlenecks that strongly slow down relaxation, with rates scaling as $\epsilon^{-2}$ for strong defects. Using large-deviation theory, we characterize rare dynamical trajectories and identify distinct relaxation pathways associated with low- and high-activity regimes in action space. We provide numerical evidence that the large-deviation function exhibits a dynamical phase transition in the limit $\epsilon \to \infty$. We then extend our analysis to the nonlinear case, considering a single nonlinear defect embedded in either a linear or a fully nonlinear discrete Schr\"odinger equation. Numerical simulations reveal a qualitatively faster approach to equilibrium than in the corresponding linear-defect model, driven by the amplitude-dependent weakening of the defect. Our results provide a unified framework for understanding thermalization, rare fluctuations, and relaxation pathways in stochastic tight-binding systems.
\end{abstract}


\maketitle

%\tableofcontents

\section{Introduction}

Thermalization, namely, the approach to equilibrium, is a classic problem in statistical mechanics and still continues to receive significant attention, both for classical and quantum dynamics.
On general grounds, it amounts to understanding how a system evolves from a non-equilibrium (possibly low-entropy) configuration towards a  state of maximal entropy, as prescribed by basic thermodynamics. 
A wide range of models have been investigated in this context. 
Among them, tight-binding chains represent one of the simplest and most widely used models, providing a minimal description of wave propagation and transport in lattice systems with nearest-neighbor coupling~\cite{ashcroft_solid_1976,grosso_solid_2000}. 
These models capture essential aspects of energy bands, dispersion, and interference effects, and form the basis for understanding phenomena ranging from electronic conduction in solids to light propagation in photonic lattices and cold atoms in optical potentials~\cite{christodoulides_discretizing_2003,morsch_dynamics_2006}. 
However, in practical implementations, ideal translational symmetry is often broken by defects, impurities, or structural disorders, which can have profound consequences for both spectral properties and dynamical behavior~\cite{lee_disordered_1985,abrahams_50_2010}. 
In particular, the celebrated phenomenon of Anderson localization, i.e., the absence of diffusive transport due to strong wave interference in disordered media, was originally predicted in a tight-binding setting with random on-site energies and has since been observed in a variety of physical contexts~\cite{anderson_absence_1958, roati2008anderson}.

Even a single impurity or defect embedded in an otherwise periodic lattice can create localized eigenstates with energies outside the band of extended states, drastically altering transport and relaxation~\cite{anderson_localized_1961,mott_theory_1961,economou_greens_2006}. 
These impurity‐induced bound states have been studied in electronic systems~\cite{lee_disordered_1985}, photonic crystals~\cite{christodoulides_discretizing_2003,joannopoulos_photonic_2011}, and engineered quantum simulators~\cite{bruderer_probing_2006,klein_dynamics_2007,palzer_quantum_2009}, where they act as scattering centers and can trap excitations locally. 
The interplay of localized and extended modes governs relaxation and equilibration processes after a local perturbation, and plays a key role in impurity models across many disciplines~\cite{john_strong_1987, hewson_kondo_1993,lepri_thermalization_2023}.

In realistic settings, lattice systems are never perfectly closed or isolated but interact with their environment, leading to decoherence and dephasing~\cite{breuer_theory_2007,zurek_decoherence_2003}. 
Open system dynamics is commonly described by using Lindblad master equations, which capture Markovian dissipation and noise while preserving complete positivity~\cite{breuer_theory_2007, lindblad_generators_1976,gorini_completely_1976, chruscinski_brief_2017, manzano_short_2020}. 
In lattice models, related non-unitary processes such as repeated projective measurements and stochastic resetting have been shown to modify survival probabilities in tight-binding chains and may give rise to localization effects depending on the measurement protocol~\cite{das_quantum_2022,das_quantum_2022-1,dattagupta_stochastic_2022}.
Local dephasing, in particular, suppresses phase coherence among sites without directly exchanging energy with the environment, and is known to convert ballistic transport into diffusive behavior even in simple tight-binding chains~\cite{ishiyama2025exact_den,ishiyama2025exact_curr}. 
The combination of defects and dephasing raises fundamental questions about how environmental noise influences the competition between localization and relaxation: 
Does dephasing destroy defect-induced localization, or can localized modes survive and hinder equilibration?
How do rare dynamical pathways contribute to the overall relaxation process?

To address these questions, we study the relaxation dynamics of a tight-binding chain subject to local dephasing noise in the presence of defects. 
We first focus on a single linear on-site defect for which analytical treatment is possible. First, we derive analytical expressions for the defective eigenvalues and eigenvectors in terms of those of the defect-free system. 
This enables us to formulate the dephasing dynamics in the eigenstate basis and to identify a doubly stochastic overlap matrix as the central object, which governs relaxation. By following a kinetic approach, we obtain a master equation for mode populations that resembles a continuous-time random walk in action space and
is equivalent to a suitable Lindblad description.

Solving this master equation reveals that relaxation towards the steady state is, as expected, exponential in time, with a rate set by the spectral gap of the overlap matrix. 
Defect-induced localized modes act as bottlenecks for dephasing-induced transport in the action space, slowing down relaxation and leading to a relaxation rate which depends on the defect strength, according to a characteristic algebraic law. 

To go beyond typical behavior and characterize rare relaxation pathways, we employ the large-deviation theory~\cite{touchette_large_2009,touchette_introduction_2018,jack_ergodicity_2020} to the stochastic dynamics in action space. 
This approach to studying thermalization~\cite{zannetti_condensation_2014} has been proposed recently 
for harmonic networks perturbed by noise~\cite{lepri_large-deviations_2024} but, to our knowledge, has not been applied so far for tight-binding models. 
Within this framework, a tilted generator is introduced that biases trajectories according to their activity (i.e. the number of state changes), allowing the identification of distinct classes of relaxation pathways dominated by either localized or extended modes.

Finally, we extend our analysis to nonlinear defects subject to dephasing noise, for which closed analytical solutions analogous to the linear case are not possible, and numerical simulations are used to explore relaxation behavior.
We first consider a single nonlinear defect embedded in an otherwise linear tight-binding chain, which provides the most direct nonlinear analogue of the linear-defect problem described above~\cite{molina_nonlinear_1993,tsironis_generalized_1994,brazhnyi_spontaneous_2011,lepri_asymmetric_2011,dambroise_eigenstates_2013}. 
We then consider the case where nonlinearity is present at all lattice sites,
leading to the discrete nonlinear Schr\"odinger (DNLS) equation~\cite{eilbeck_discrete_1985, rasmussen_statistical_2000, kevrekidis_discrete_2009}.
The DNLS equation has a long history as a paradigmatic model for nonlinear lattice dynamics and has provided a fundamental framework for studying discrete breathers and intrinsic localized modes~\cite{flach_discrete_1998, flach_discrete_2008}. 
It has also been extensively investigated in connection with breather decay~\cite{johansson_dynamics_1998,iubini_discrete_2013}, metastability~\cite{iubini_effective_2025}, adiabatic invariants~\cite{politi_frozen_2022}, and long-time relaxation and thermalization processes~\cite{flach_discrete_2008,iubini_dynamical_2019}. 

The relaxation and stability of localized states in the DNLS equation have also been explored through various stochastic frameworks. Early investigations established that thermal fluctuations limit the lifetime of breather-like excitations~\cite{rasmussen_localized_1998}. Subsequent research utilized simplified stochastic models, such as the microcanonical Monte Carlo and partial exclusion process, to characterize the coarsening dynamics of breathers and their weak coupling to the background~\cite{iubini_discrete_2013, iubini_coarsening_2014, iubini_relaxation_2017}. 
More recently, the focus has shifted toward multiplicative dephasing noise, which, unlike additive noise, preserves the system's norm while allowing energy exchanges~\cite{ebrahimi_dynamics_2025,ebrahimi_stochastic_2025}.

The goal of the present work is not to address the broad DNLS relaxation problem, but rather to use the nonlinear models as natural extensions of the defect problem and to investigate how the dephasing-induced relaxation mechanisms identified for linear defects are modified when the defect becomes amplitude dependent.
We find that nonlinear defects lead to a qualitatively different relaxation dynamics compared to the linear case,
including a faster relaxation which results from the amplitude-dependent reduction of the effective defect strength.
Yet, some properties of the linear defect model turn out to be valid also in the nonlinear case in terms of an effective description restricted to times much shorter than typical relaxation time scales.
Together, our results provide a comprehensive and unified picture of how defects, dephasing noise, and dynamical fluctuations shape relaxation and thermalization in noisy lattice systems.

The paper is organized as follows.
In Section~\ref{sec:linDefect}, we introduce the tight-binding chain with a single linear defect and discuss its spectral properties.
Section~\ref{sec:dephasing} is devoted to the dephasing dynamics consisting of stochastic phase kicks. 
We derive approximate kinetic equations, consistent with those obtained via the usual Lindblad formalism. 
Such equations are linear and allow one to analyze the steady state and relaxation time through the spectral analysis.
In Section~\ref{sec:largeD}, we extend the analysis using a large-deviation approach, which allows us to characterize rare relaxation pathways and activity fluctuations in action space.
Section~\ref{sec:nonlinDefects} addresses nonlinear defects, starting from a single nonlinear defect and then considering the discrete nonlinear Schr\"odinger equation and generalized nonlinearities.
Finally, Section~\ref{sec:conclu} summarizes our results with possible extensions, while some technical details and complementary derivations are provided in the Appendices.

%================================================
%================================================
%================================================
\section{Tight-binding chain with linear defect}
\label{sec:linDefect}

We consider a tight-binding chain with $N$ sites and periodic boundary conditions in the presence of an on-site potential of strength $\epsilon$ at site $M$ representing a defect. 
The  Hamiltonian of the system is given by
\begin{equation}
\label{eq:H_defect}
    {H} = - C \sum_{j=0}^{N-1} ( \ket{j}\bra{j+1} + \ket{j+1}\bra{j} ) - \epsilon \ket{M} \bra{M},
\end{equation}
where $C > 0$ is the hopping amplitude and the lattice site basis is set by the Wannier state $\ket{j}$ such that the probability amplitude of finding a particle with state $\ket{\psi}$ at site $j$ is given by the projection $\psi_j \equiv \braket{j}{\psi}$.  
The periodic boundary condition $\ket{N}=\ket{0}$ renders all lattice sites equivalent, so that the location of the defect can be chosen without loss of generality.
The Hamiltonian may also be written as $H = H_{\mathrm{TBC}} - \epsilon \ket{M} \bra{M} $, where $H_{\mathrm{TBC}} = -C \sum_{j=0}^{N-1} ( \ket{j}\bra{j+1} + \ket{j+1}\bra{j} )$ is the defect-free tight-binding Hamiltonian. 
Setting $\hbar = 1$, the unitary evolution of the system is described by the time-dependent Schr\"odinger equation~\cite{economou_greens_2006,acharya_dynamics_2026}
\begin{align} 
\label{eq:dls+def}
\ii \dv{\psi_j}{t} = -C  ( \psi_{j+1}+\psi_{j-1}  ) - \epsilon \,\delta_{jM}\,\psi_M,
\end{align}
with the boundary condition $\psi_{j+N}(t)=\psi_j(t)$ and $\delta_{jM}$ denoting the Kronecker delta. 
Here, the defect is referred to as linear, as it contributes a term proportional to the local amplitude $\psi_M$ in the Schr\"odinger equation.
The expected value of the total energy  $E= \langle \psi | H | \psi \rangle$ of the system reads
\begin{align}
E = -C \sum_{j=0}^{N-1} \big( \psi_j^{*}\psi_{j+1} + \psi_{j+1}^{*}\psi_j \big) - h_\epsilon ,
\end{align}
where $h_\epsilon=\epsilon\,|\psi_M|^2$ denotes the negative of the defect energy.

The eigenvalues $\omega_\nu$ and eigenvectors $\ket{\chi^\nu }$ of the defect-free Hamiltonian $H_{\mathrm{TBC}}$, which satisfy the relation 
$H_{\mathrm{TBC}} \ket{\chi^\nu } = \omega_\nu \ket{\chi^\nu }$, 
are given by
\begin{align}
    \omega_\nu = -2 C  \cos(2\pi \nu/N),  \quad \mbox{and}\quad 
    \chi^\nu_n = \frac{1}{\sqrt{N}} e^{\ii 2 \pi \nu n / N } ,
    \label{eq:specrum-freeH}
\end{align}
respectively, where $ \nu = 0,1,\ldots,N-1$ and  $\chi^\nu_n = \braket*{n}{\chi^\nu }$.
As we have $\omega_{N-\nu} = \omega_\nu$, the pairs of eigenvectors $\ket{\chi^\nu }$ and $\ket*{\chi^{N-\nu} }$ form doubly--degenerate subspaces. Table~\ref{tab:tbm} summarizes the possible scenarios for both odd and even $N$.
\begin{table*}[t]
\centering
\begin{tabular}{p{0.7cm}@{\hspace{0.7cm}}p{4.4cm}@{\hspace{0.7cm}}p{2.5cm}@{\hspace{0.7cm}}p{2.7cm}}
\hline\hline
$N$ & Number of doubly--degenerate states (each pair of $\nu$ and $N-\nu$) 
    & Number of non--degenerate states
    & Number of distinct energy eigenvalues \\[3pt]
\hline\hline
even & $(N-2)/2$ 
     & $2$ (for $\nu=0,N/2$)
     & $N/2+1$ \\[0.5ex]
odd  & $(N-1)/2$ 
     & $1$ (for $\nu=0$)
     & $(N+1)/2$ \\
\hline
\end{tabular}
\caption{Degeneracy of $H_{\mathrm{TBC}}$ on a chain of size $N$ with $ \nu = 0,1, \ldots, N-1$.}
\label{tab:tbm}
\end{table*}
The defect-free spectrum in Eq.~\eqref{eq:specrum-freeH} forms a band bounded by
$E=\pm 2C$, corresponding to the standard tight-binding dispersion~\cite{ashcroft_solid_1976}.

\begin{figure}[t]
    \centering
    \includegraphics[scale=0.89]{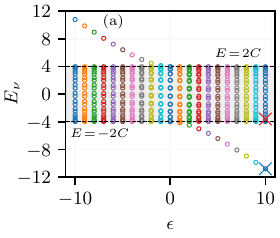}\hskip-1pt
    \includegraphics[scale=0.89]{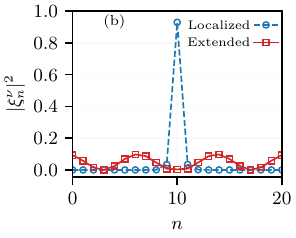}
    \caption{Spectrum of the defective Hamiltonian~\eqref{eq:H_defect} for a chain of size $N=21$ with a defect located at site $M=10$, and $C=2$. (a) Eigenvalues $E_\nu$ denoted by circles as a function of the defect strength $\epsilon$. The gray region denotes the energy band, with dashed black lines marking the band edges at $E=\pm 2C$. (b) Spatial distribution $|\xi_n^\nu|^2 \equiv |\braket*{n}{\xi^\nu}|^2$ of two representative eigenstates $\ket*{\xi^\nu}$ at $\epsilon=10$, marked by the blue ($E_\nu \approx -10.77 $) and red ($E_\nu \approx -3.95 $) crosses on panel (a). The blue curve (corresponding to the blue cross) depicts the defect-localized mode, while the red curve (corresponding to the red cross) represents an extended mode.}
    \label{fig:H-defect-spectrum}
\end{figure}

Let us denote the eigenvalues of the defective Hamiltonian $H$ by  $E_\nu$ and eigenvectors $\ket{\xi^\nu }$, respectively, such that one has
$H \ket{\xi^\nu } = E_\nu \ket{\xi^\nu }$.
Introducing a single on-site defect breaks the degeneracy of the defect-free spectrum and generates exactly one defect-induced bound state
whose eigenvalue detaches from the band $[-2C , 2C]$, see Fig~\ref{fig:H-defect-spectrum}(a).
For instance, it may be seen in the figure, this bound-state eigenvalue lies far below the lower band edge $-2C$ for large enough $\epsilon>0$.
On the other hand, the remaining $N-1$ eigenvalues stay either within the band or very close to the band and correspond
to extended states that are only weakly affected by the defect.

For the representative case $\epsilon=10$, Fig.~\ref{fig:H-defect-spectrum}(a) highlights the bound-state eigenvalue outside the band and a typical in-band eigenvalue, marked by the blue and the red cross, respectively.
The corresponding eigenstates  are shown in Fig.~\ref{fig:H-defect-spectrum}(b): the state outside of the band 
is exponentially localized around the defect site, with its probability density
sharply peaked at $n=M$ and rapidly decaying away from it.
Conversely, the state within the band remains spatially extended with probability distributed across the lattice.
This separation between a defect-localized mode and extended band modes forms
the spectral basis for the distinct relaxation behaviors discussed below.

The quantities $E_\nu$ and $\ket*{\xi^\nu}$ can be obtained analytically
from the eigenvalues $\omega_\nu$ and eigenvectors $\ket*{\chi^\nu}$ of the
defect-free Hamiltonian.
In the following, we focus on odd $N$ and present the analytical construction in Appendix~\ref{sec:app-defect-free-H}, while the case of even $N$
can be treated in an analogous manner.
The introduction of a defect breaks the degeneracy of $H_{\mathrm{TBC}}$, such that $(N-1)/2$ eigenvalues and corresponding eigenvectors of $H$ are given by [see Appendix~\ref{sec:app-defect-free-H}]
\begin{align}
    E_\nu = \omega_\nu, \quad \mbox{with} \quad 
    \ket{\xi^\nu } =  \frac{1}{\sqrt{2}}\qty[ \ket{\chi^\nu } - \frac{\chi^\nu_M}{(\chi^\nu_M)^*} \ket*{\chi^{N-\nu} } ] ,
    \label{eq:H-eigval-first-set}
\end{align}
for $ \nu = 1, 2, \ldots, (N-1)/2$.
The remaining $(N+1)/2$ eigenvalues of $H$ are given by the roots of $(N+1)/2$--degree secular equation 
\begin{align}
    \frac{1}{E_\mu-\omega_0} + \sum_{\nu=1}^{(N-1)/2} \frac{2}{E_\mu-\omega_\nu } + \frac{N}{\epsilon} =0 ,
    \label{eq:secular1}
\end{align}
with $\mu \in [0, (N+1)/2, (N+3)/2 ,\ldots (N-1)]$, and the corresponding eigenvectors read
\begin{align}
\label{eq:H-eigenvec-remain}
\ket{\xi^\mu} = \frac{1}{{\mathcal N}_\mu} \sum_{\lambda=0}^{N-1}  \dfrac{ (\chi_M^{\lambda})^*} {\omega_\lambda-E_\mu}\, \ket*{\chi^\lambda} ,
\end{align}
where ${\mathcal N}_\mu \equiv  (  \sum_{\lambda=0}^{N-1} (\omega_\lambda-E_\mu)^{-2} / N )^{1/2}   $.

Under the unitary evolution governed by Eq.~\eqref{eq:dls+def}, the local occupation probability $P_n(t)\equiv|\psi_n(t)|^2$ displays temporal oscillations in the defect-free case $\epsilon=0$~\cite{das_quantum_2022}, whereas a finite defect strength $\epsilon$ perturbs this coherent dynamics nonlinearly, resulting in a nontrivial modification of the oscillatory behavior~\cite{acharya_dynamics_2026}.
In this work, we focus, instead, on the relaxation under the  stochastic dynamics induced by dephasing noise. The resulting nonunitary evolution, which suppresses quantum coherence and qualitatively alters the long-time behavior, is examined in the section below. 

%================================================
%================================================
%================================================
\section{Dynamics with dephasing}
\label{sec:dephasing}

To include the effect of environmental decoherence while conserving particle number, we introduce a local, pure dephasing noise acting on the lattice sites.
The stochastic time evolution is implemented in continuous time as follows.
Starting from time $t$, the system undergoes a unitary evolution under the defective Hamiltonian $H$ for a random waiting time $\tau$, during which the wave function evolves as $\psi(t) \to \psi(t+\tau)=\exp(- \ii H\tau) \psi(t)$.
The waiting time $\tau$ between successive noise events is sampled independently from an exponential distribution $p(\tau)=\beta\,\mathrm{e}^{-\beta \tau}$, which defines the average rate $\beta$ of dephasing events.
At the end of this interval, at time $t+\tau$, a lattice site $n$ is selected uniformly at random and the corresponding wave-function amplitude acquires a stochastic phase, $\psi_n(t+\tau) \to \mathrm{e}^{\ii \theta_n} \psi_n(t+\tau)$, where the phase $\theta_n$ is drawn independently from a prescribed distribution $g(\theta)$.
This procedure is then repeated, generating a piecewise unitary time evolution interrupted by random local dephasing events.

Local dephasing in the site basis has a clear physical meaning: 
it does not directly change the probability of finding the particle on a given site, but it induces random, site-dependent, local phase fluctuations acting on the quantum state. 
As a result, relative phase information between different lattice sites is progressively lost, leading to the decay of phase coherence between amplitudes on different sites and the suppression of quantum interference effects.

\subsection{Kinetic approach}
\label{ssec:kin}

While the microscopic dynamics is stochastic at the level of individual trajectories, the ensemble-averaged evolution is conveniently described in terms of a kinetic approach as follows.

To study the relaxation dynamics, we switch to the normal-mode basis $\{ \ket*{\xi^{\nu}} \}$ that diagonalizes the Hamiltonian $H$.
The mode amplitudes in this basis are defined as $a_\nu=\langle\xi^{\nu}|\psi\rangle$.
We write
\begin{align}
a_\nu \equiv \sqrt{I_\nu}\,\mathrm{e}^{\ii \phi_\nu},
\label{eq:Inu-def}
\end{align}
which defines $I_\nu = |a_\nu|^2$ as the population of the normal mode $\ket*{\xi^{\nu}}$ with phase $\phi_\nu$.

Since a local dephasing event at site $n$ multiplies the amplitude $\psi_n$ by $e^{\ii \theta_n}$ while leaving all other components unchanged, it can be represented by a diagonal unitary operator $\widetilde D^{(n)}(\theta_n) = \mathbb{I} + \big(e^{\ii \theta_n} - 1\big) |n\rangle \langle n| $, which in the normal-mode basis reads
\begin{align}
D^{(n)}_{\nu\mu}(\theta_n) = \delta_{\nu\mu} + \big(e^{\ii \theta_n} - 1\big)\xi_n^{\nu *} \xi_n^{\mu}.
\end{align}
Acting with the operator on the mode amplitudes, one obtains $a_\nu' = \sum_\mu D^{(n)}_{\nu\mu} a_\mu$, or in matrix form
\begin{align}
& a'  = (\mathbb{I} +  U^{(n)}) a, \label{eq:a-mat}\\
& U^{(n)}_{\nu\mu} = \big(e^{\ii \theta_n} - 1\big)\xi_n^{\nu *} \xi_n^{\mu} ,
\label{eq:Unumu}
\end{align}
such that the operator $U^{(n)}$ mixes mode population according to their overlap at site $n$.

As under unitary evolution the mode amplitude evolve as $a_\nu(t) = a_\nu(0) \exp(-\ii E_\nu t) $, 
in a random dephasing collision interval $\tau$ from $t$ to $t+\tau$, we may write from Eq.~\eqref{eq:a-mat} that
\begin{align}
    a(t+\tau) = (\mathbb{I}  +  U^{(n)}) e^{- \ii \Omega \tau} a(t) ,
    \label{eq:modedyn}
\end{align}
where $\Omega = \text{diag}(E_0, \dots, E_{N-1})$.
Immediately after a dephasing event, the mode amplitude updates as $a_\nu' = a_\nu + \Delta a_\nu$, with $\Delta a_\nu = \sum_\mu U^{(n)}_{\nu\mu} a_\mu$, and the corresponding population becomes $I_\nu' = |a_\nu'|^2 = (a_\nu^* + \Delta a_\nu^*)(a_\nu + \Delta a_\nu)$, which gives the change $ \Delta I_\nu = I_\nu' - I_\nu$ to the leading order as
\begin{align}
    \Delta I_\nu &\!=\! \sum_{\mu} \qty( U^{(n)}_{\nu\mu}  \sqrt{ I_\nu I_\mu }  e^{\ii (\phi_\mu - \phi_\nu)} 
\!+\! U^{(n)*}_{\nu\mu} \sqrt{ I_\nu I_\mu }  e^{\ii (\phi_\nu - \phi_\mu)} ) \nonumber \\
&+ \sum_{\mu\mu'}  U^{(n)*}_{\nu\mu'}  \, U^{(n)}_{\nu\mu}  \sqrt{ I_\mu I_\mu' }  e^{\ii (\phi_\mu - \phi_\mu')} .
\end{align}
Invoking the kinetic random-phase approximation and assuming the phases are uniformly distributed and uncorrelated,
$\langle e^{\ii (\phi_\mu-\phi_\nu)} \rangle = \delta_{\mu\nu}$, yield the phase-averaged change $\Delta \overline{I}_\nu = (  U^{(n)}_{\nu\nu} + U^{(n)*}_{\nu\nu} )  \overline{I}_\nu + \sum_{\mu} | U^{(n)}_{\nu\mu} |^2\,\overline{I}_\mu $,
which along with Eq.~\eqref{eq:Unumu}, reduces to
\begin{align}
& \Delta \overline{I}_\nu
 = \sum_{\mu} K^{(n)}_{\nu \mu} \overline{I}_\mu, 
 \label{eq:Inu-avg-k}\\
& K^{(n)}_{\nu \mu} = 4\sin^2\! \qty(\frac{\theta_n}{2} )  \Big[|\xi_n^{\nu}|^2|\xi_n^{\mu}|^2  -\,|\xi_n^{\nu}|^2\,\delta_{\nu\mu}  \Big] ,
\label{eq:Inu-phase-averaged}
\end{align}
Equation~\eqref{eq:Inu-avg-k} and~\eqref{eq:Inu-phase-averaged} represent the single-collision kinetic update of the mode population $\overline{I}_\nu$ after averaging over random phases.

For $m$ independent collisions within the time interval $[t,t+T]$, the population evolves as per Eq.~\eqref{eq:Inu-avg-k} as $\overline I(t+T)= \prod_{k=1}^m (\mathbb I+K^{(n_k)} ) \overline I(t)$, where the product is time-ordered.
For uniformly chosen dephasing sites $n_k$, Poisson-distributed waiting time distribution $p(\tau)$ with mean $\langle\tau\rangle$, and many weak, uncorrelated collisions, we approximate $\prod_{k=1}^{m} (1 + K^{(n_k)}) \approx  \exp  (\sum_{k=1}^{m} K^{(n_k)} )$.
The average number of collisions during the interval being $ {T}/{\langle \tau \rangle}$, leads to the relation
\begin{align}
\overline{I}(t+T)
\simeq
\exp\!\left(\frac{T}{\langle\tau\rangle}\,
\mathbb{E}_{n,\theta}[K^{(n)}]\right)\overline{I}(t) , \label{eq:Iavgttau}
\end{align}
where $\mathbb{E}_{n,\theta}$ denotes the remaining ensemble averaging over $n$ and $\theta$.
Averaging over the distribution $g(\theta)$, defining
\begin{align}
\kappa \equiv \expval{ 4\sin^2({\theta}/{2})}_{g(\theta)} ,
\label{eq:kappa-def}
\end{align}
and considering uniform distribution for dephasing sites $n$, we obtain from Eq.~\eqref{eq:Inu-phase-averaged} that $\mathbb{E}_{n,\theta}[K^{(n)}_{\nu\mu}] = ({\kappa}/{N}) [ \sum_{n} |\xi_n^\nu|^2|\xi_n^{\mu}|^2 - \delta_{\nu\mu} ]$, 
which when substituted in Eq.~\eqref{eq:Iavgttau} produces in the continuous limit ($T \to 0$) the differential form
\begin{align}
\dot{\overline{I}}_\nu &= \frac{\kappa \beta }{N }\, \sum_{\mu} \Big[ \sum_{n}|\xi_n^\nu|^2|\xi_n^{\mu}|^2 - \delta_{\nu\mu}  \Big]\,\overline{I}_{\mu},
\label{eq:Tbar_dot-oo}
\end{align}
where we have used $\langle\tau\rangle = 1/\beta$ for the exponential distribution $p(\tau)$.
For a uniformly distributed phase $\theta\in[0,2\pi)$, one has $g(\theta)=1/(2\pi)$ and therefore $\kappa=2$, which is the case considered throughout this work.
Upon identifying the dephasing rate as 
\begin{align}
\gamma=\kappa\beta/N ,
\end{align}
it may be recast in matrix form as
\begin{align}
   \dot{\overline{I}} = \gamma (W - \mathbb{I}) {\overline{I}}, 
   \label{eq:Idot}
\end{align}
where the overlap matrix $W$ is defined by its elements
\begin{align}
   \qquad W_{\nu\mu} \equiv \sum_{n} |\xi_n^\nu|^2\,|\xi_n^\mu|^2 .
   \label{eq:W-def}
\end{align}
We note that the matrix $W$ is real, symmetric, and~\emph{doubly stochastic}~\cite{perfect_spectral_1965}, i.e., 
$\sum_{\mu} W_{\nu\mu} = \sum_{\nu} W_{\nu\mu} = 1$. 
Moreover, the structure of $W$ reflects the spatial character of the eigenstates and controls how dephasing redistributes population between them.
Note that Eq.~\eqref{eq:Idot} remains valid for the dephasing dynamics under the defect-free ($\epsilon =0$) tight-binding Hamiltonian $H_{\mathrm{TBC}}$, with only modification in the definition of $W$, which in this case reduces to  $W_{\nu\mu} = \sum_{n} |\chi_n^\nu|^2\,|\chi_n^\mu|^2 = 1/N$ [see Eq.~\eqref{eq:specrum-freeH}], implying that local dephasing mixes all modes uniformly.

When a macroscopic fraction $f$ of the $N$ sites is dephased during each stochastic event, the structure of the effective kinetic equation for the mode populations Eq.~\eqref{eq:Idot} remains unchanged.
Each local phase randomization produces the same averaged redistribution between eigenstates, determined by their spatial overlaps encoded in the matrix $W$.
Dephasing multiple sites in a single event therefore corresponds to applying several independent phase kicks within the same time interval.
In the Markov limit, these independent contributions add up and lead to a renormalized dephasing rate $\gamma_{\mathrm{eff}} = f N \gamma = \kappa \beta f$, while the mixing matrix $W$, the steady state and the relaxation modes remain unaffected.
Consequently, increasing the fraction of dephased sites accelerates relaxation without altering the underlying mode--coupling mechanism.

\subsection{Comparison with Lindblad formalism}

We now analyze the dephasing dynamics using the density operator $\rho$. 
Such a dynamics governing the evolution of the tight-binding chain can be formulated, for the case of a linear defect, within the Lindblad framework, which provides the most general form of a generator for open quantum dynamics~\cite{breuer_theory_2007, lindblad_generators_1976,gorini_completely_1976, chruscinski_brief_2017, manzano_short_2020}. 
In this section, we first specialize the Lindblad equation to pure dephasing acting on a single lattice site and derive the resulting evolution equation for the mode populations.
The local dephasing is described by~\cite{schwarzer_moments_1972}
\begin{equation}
    \dot{\rho} = - \ii [H,\rho] + \sum_{j=0}^{N-1} \qty( L_j \rho L_j^\dagger - \frac{1}{2}\{L_j^\dagger L_j,\rho\}  ),
    \label{eq:lindblad-general-app}
\end{equation}
where the Lindblad operator $L_j$ reads
\begin{equation}
    L_j = \sqrt{\gamma} \ket{j} \! \bra{j},
    \label{eq:L_n-def}
\end{equation}
with a dephasing rate $\gamma > 0$.
The first term on the right-hand-side of Eq.~\eqref{eq:lindblad-general-app} corresponds to the unitary evolution, while the second term acts as the Lindblad dissipator due to dephasing. 
Substituting Eq.~\eqref{eq:L_n-def} in Eq.~\eqref{eq:lindblad-general-app}, we obtain
\begin{equation}
    \dot{\rho} = -\ii [H,\rho] + \mathcal{D}_1[\rho] +  \mathcal{D}_2[\rho].
    \label{eq:lindblad-site-app}
\end{equation}
where the dissipating terms are given by
\begin{align}
    \mathcal{D}_1[\rho]
    &= \gamma\sum_n \ket{n} \! \mel{n}{\rho}{n} \bra{n}, \label{eq:diss-def1} \\
    \mathcal{D}_2[\rho]
    &= -\frac{\gamma}{2}\sum_n \{\ket{n} \! \bra{n},\rho\} . \label{eq:diss-def2}
\end{align}

Denoting the matrix elements of the density operator in the site basis as  $\rho_{nm} = \mel{n}{\rho}{m}$, we obtain from Eqs.~\eqref{eq:lindblad-site-app}--\eqref{eq:diss-def2} that
\begin{align}
    \dot{\rho}_{nm} &= -\ii (H\rho - \rho H)_{nm} - \gamma(1-\delta_{nm})\,\rho_{nm}.
    \label{eq:rho-site-eq-app}
\end{align}
For $n=m$, we then have  $\dot{\rho}_{nn}=-\ii (H\rho-\rho H)_{nn}$, indicating that the dephasing term does not act directly on site populations, whose dynamics is governed solely by the coherent Hamiltonian evolution.
By contrast, for $n\neq m$, one finds $\dot{\rho}_{nm}=-\ii (H\rho-\rho H)_{nm}-\gamma\,\rho_{nm}$, showing that the off-diagonal elements of the density matrix decay exponentially at rate $\gamma$ due to dephasing, in agreement with the coherence loss in the site basis.

In the density-operator description, the corresponding mode population reads
$I_\nu(t)=\rho_{\nu\nu}(t)$, where
$\rho_{\mu\nu}=\langle\xi^{\mu}|\rho|\xi^{\nu}\rangle$.
To obtain the evolution equation for the mode population $I_\nu$, one expresses Eq.~\eqref{eq:lindblad-site-app} in the mode basis.
The unitary Hamiltonian part does not contribute to the time evolution of $I_\nu$ due to the fact that $\bra{\xi^\nu}[-\ii H,\rho]\ket{\xi^\nu} = 0$.
Using the mode function $\xi_n^{\nu} = \langle n | \xi^{\nu} \rangle$, one obtains
$\ket{n} \! \bra{n}  = \sum_{\mu,\nu} \xi_n^{\mu *} \xi_n^{\nu} \ket{\xi^\mu}\bra{\xi^\nu}$, and computes from Eq.~\eqref{eq:lindblad-site-app} that 
\begin{align}
 \dot{I}_\nu 
&= \gamma\sum_n |\xi_n^\nu|^2 \sum_{\mu,\sigma} \xi_n^{\mu}\,\xi_n^{\sigma *}\,\rho_{\mu\sigma}  -\frac{\gamma}{2}\sum_{n,\mu} \xi_n^{\nu *}\,\xi_n^{\mu}\,\rho_{\mu\nu} \nonumber \\
&-\frac{\gamma}{2}\sum_{n,\mu} \rho_{\nu\mu}\,\xi_n^{\mu *}\,\xi_n^{\nu}.
\label{eq:D2-general-app}
\end{align}

Let us now invoke a standard approximation in the presence of dephasing: we assume that the density matrix in the mode basis is approximately diagonal on the coarse-grained time scales $t\gg 1/\gamma$.
Hence, we may write $\rho(t) \approx \sum_\mu I_\mu(t)\,\ket{\xi^\mu}\bra{\xi^\mu}$ and  $\rho_{\mu\sigma}(t) \approx I_\mu(t)\,\delta_{\mu\sigma}$, using which 
along with $\sum_{n} |\xi_n^{\nu}|^2 = 1$, one obtains from Eq.~\eqref{eq:D2-general-app} that
\begin{align}
    \dot{I}_\nu &= \gamma \sum_{\mu} \Big[   \sum_{n} |\xi^{\nu}_n|^2 |\xi^{\mu}_n|^2 - \delta_{\nu \mu} \Big] I_\mu ,
    \label{eq:T_dot-Lindblad}
\end{align}
which coincides with Eq.~\eqref{eq:Tbar_dot-oo} as expected in this limit.

Although both formulations lead to the same effective dynamics of the mode populations, the kinetic description offers a complementary physical interpretation by explicitly revealing the microscopic sources of stochasticity underlying dephasing. 
This dual perspective establishes a transparent connection between the Lindblad framework and a classical rate-equation description.

\subsection{Steady state}

Under dephasing, the relaxation dynamics reduces to the linear Markov master equation~\eqref{eq:Idot}, governed by the spectrum of the matrix $W$.
Since $W$ is doubly stochastic, the Perron--Frobenius theorem ensures that its largest eigenvalue is $\lambda_1=1$, with the corresponding (right) eigenvector $u=(1,1,\ldots,1)^\top$, and all other eigenvalues satisfy $-1 \le \lambda_{2\leq i \leq N} \le  1$~\cite{perfect_spectral_1965}.
Moreover, $W$ being positive semi-definite restricts its eigenvalues to $0\leq \lambda_i \leq 1$.
From Eq.~\eqref{eq:Idot}, one obtains 
$\frac{\mathrm d}{ \mathrm{d} t}(u^\top\overline I)=\gamma \,u^\top(W-\mathbb I)\overline I=0$, suggesting that the total norm or mode population $A \equiv \sum_\nu\overline I_\nu = u^\top\overline I$ is conserved.
Consequently, the steady state $\overline I^\infty$, which satisfies $W\overline I^\infty= \overline I^\infty$, uniquely yields $\overline I^\infty = (A/N)u$.
For a normalized wave function, $A=\sum_\nu |a_\nu|^2=\langle\psi|\psi\rangle=1$, so that $\overline I_\nu^\infty=1/N$ for all modes.
The steady-state energy is therefore given by the uniform spectral average, $E(\infty) \equiv \langle H\rangle (t\to\infty)=\sum_\nu E_\nu \overline I_\nu^\infty =\mathrm{tr}(H)/N$.
Since the hopping terms are off-diagonal in the site basis and the defect contributes $-\epsilon$ to the trace, one has $\mathrm{tr}(H)=-\epsilon$, yielding $E(\infty)=-\epsilon/N$, independent of the initial state.
On the other hand, for $\epsilon = 0$, i.e., for dephasing under defect-free $H_{\mathrm{TBC}}$, one has $\mathrm{tr}(H)=0$, and hence the average total energy in the steady-state vanishes.

The corresponding spatial distribution follows directly from the mode expansion $\psi_n(t)=\sum_\nu a_\nu(t)\,\xi_n^{\nu}$.
Under dephasing, the phases of the mode amplitudes become random and uncorrelated, so phase averaging removes interference terms and gives the occupation probability $P_n(t) \equiv \langle|\psi_n(t)|^2\rangle=\sum_\nu \overline I_\nu(t)\,|\xi_n^{(\nu)}|^2$.
With uniform steady-state mode population $\overline I_\nu^\infty=1/N$, the occupation probability, in both defective ($\epsilon \neq 0$) and defect-free ($\epsilon =0$) cases, becomes uniform as well, $P_n(t\to\infty)=1/N$, indicating complete loss of memory of any initial spatial localization.

We note that in the context of wave systems, the uniform steady state in mode space corresponds to the equilibrium described by the infinite temperature limit of a Rayleigh–Jeans distribution~\cite{nazarenko_wave_2011,ebrahimi_dynamics_2025,hannani_stochastic_2023}.
Dephasing thus drives the system towards a maximal-entropy state constrained solely by norm conservation, resulting in uniform populations in both mode and real space.

\subsection{Relaxation rates and spectral properties}

\begin{figure}[t]
    \centering
    \includegraphics[scale=0.89]{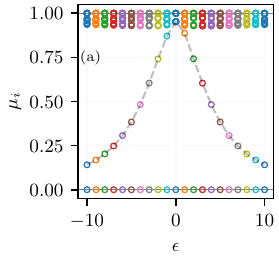} 
    \includegraphics[scale=0.89]{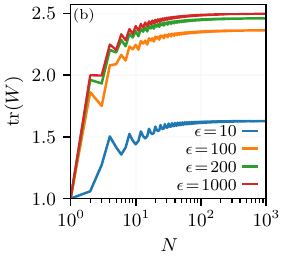}
    \caption{(a) Dependence of the rates $\mu_i$ on the defect strength $\epsilon$. The points are obtained from Eq.~\eqref{eq:mui-def} with $\gamma=1$, using numerically computed eigenvalues $\lambda_i$ of the overlap matrix $W$ for $N=21$ and $C=2$. The gray dashed line highlights the relaxation rate $\mu_2$. (b) Trace of $W$ as a function of the lattice size $N$ for different defect strengths $\epsilon$, with $C=20$. In both panels, the matrix $W$ is constructed from Eq.~\eqref{eq:W-def} using numerically obtained eigenvectors $\ket*{\xi^\nu}$ of the defective Hamiltonian~\eqref{eq:H_defect}.}
    \label{fig:tracW}
\end{figure}

The symmetry of the overlap matrix $W$ guarantees the existence of an orthonormal eigenbasis $ \{ v_i \}_{i=1}^N$, allowing one to write $W = V \Lambda V^\top$, where $V = [v_1, v_2, \ldots, v_N]$ is the orthogonal matrix whose columns are the normalized eigenvectors of $W$, and $\Lambda=\mathrm{diag}(\lambda_1,\ldots,\lambda_N)$ collects the corresponding eigenvalues. 
The leading eigenpair is given by $\lambda_1=1$ with eigenvector $v_1 = u/\sqrt{N}$.
Expanding the initial condition as $\overline I(0)=\sum_i c_i v_i$, the solution of Eq.~\eqref{eq:Idot} reads
\begin{align}
\overline I(t)=\sum_{i=1}^N c_i\,e^{- \mu_i t}\,v_i ,
\label{eq:Isol}
\end{align}
where the rates $\mu_i$ read
\begin{align}
    \mu_i \equiv \gamma (1-\lambda_i).
    \label{eq:mui-def}
\end{align}
As $\mu_1=0$, the uniform component ($i=1$) of the solution~\eqref{eq:Isol} remains constant, while other components decay exponentially. 
The slowest decay, which sets the overall relaxation time $\tau_{\mathrm{relax}}$, is controlled by $\mu_2$, so that $\tau_{\mathrm{relax}} = \mu_2^{-1}= \gamma^{-1} (1-\lambda_2)^{-1}$.  
Figure~\ref{fig:tracW}(a) displays the rates $\mu_i$ as a function of the strength of the defect  $\epsilon$; in particular, the dominant relaxation rate $\mu_2$, shown by the gray dashed line, decreases symmetrically with increasing $|\epsilon|$, indicating a slower relaxation for both positive and negative defects.

There exists a bound on the relaxation time set by the trace of $W$.
Since we have $1=\lambda_1 \ge \lambda_2\ge\cdots\ge\lambda_N\ge0$, the trace identity $\sum_{i=2}^N\lambda_i=\mathrm{tr}(W)-1$ implies $\lambda_2 \le \min\{1,\mathrm{tr}(W)-1\}$.
Moreover, as $\lambda_2$ is the largest among the remaining $N-1$ eigenvalues, it is bounded from below by their average, which along with the upper bound gives $\frac{\mathrm{tr}(W)-1}{N-1} \le \lambda_2 \le \min\{1,\mathrm{tr}(W)-1\}$, which yields
\begin{align}
\max\{0,\,2-\mathrm{tr}(W)\}
\le  1-\lambda_2 
\le \frac{N - \mathrm{tr}(W)}{N-1} .
\label{eq:tau-rel-bound}
\end{align}
In the absence of defect ($\epsilon= 0$), one has $W_{\mu\nu} = 1/N$ with $\lambda_1=1$, $\lambda_{i\neq1}=0$, and $\mathrm{tr}(W)=1$, so that $\mu_2 = \tau^{-1}_{\mathrm{relax}}=\gamma$, consistent with both the lower and upper bounds in Eq.~\eqref{eq:tau-rel-bound}. 
For nonzero $\epsilon$ in a finite-sized chain, one has  $1 < \mathrm{tr}(W) < N$ since not all eigenvalues $\lambda_{i\neq1}$ vanish, see Fig.~\ref{fig:tracW}(b). Consequently, the upper bound in Eq.~\eqref{eq:tau-rel-bound} is strictly less than unity and approaches one in the limit of large $N$, as inferred from the figure. 
This implies the  rate $\mu_2 < \gamma$ and indicates that relaxation under dephasing is fastest in the absence of a defect.
On the other hand, the lower bound in Eq.~\eqref{eq:tau-rel-bound} can provide nontrivial information for sufficiently weak defect strengths, for which $\mathrm{tr}(W)<2$ even at large system sizes $N$. 
However, for stronger defect strengths, $\mathrm{tr}(W)$ exceeds $2$ already for modest system sizes, so that the lower bound reduces to the trivial condition $\mu_2 \geq 0$, except at very small $N$.

\begin{figure}[t]
\centering
\includegraphics[scale=0.92]{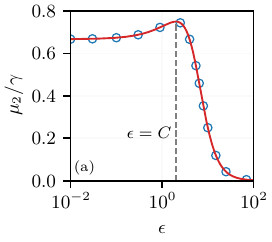} \hskip-5pt
\includegraphics[scale=0.92]{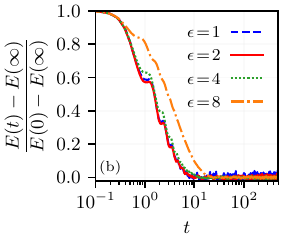}
\caption{Relaxation in trimer system with $N=3$. (a) Scaled relaxation rate $\mu_2$ as a function of $\epsilon$ showing a maximum value 0.75 at $\epsilon=C=2$. The line is obtained using Eq.~\eqref{eq:3lev-tau} with $\mu_2 = \tau^{-1}_{\mathrm{relax}}$, while the points denote $(1-\lambda_2)$, where $\lambda_2$ is obtained by numerically computing the eigenvalues of $W$.  (b) Evolution of scaled total energy as a function of time $t$ for different values of $\epsilon$ and $C=2$. The lines are obtained from $2^{15}$ stochastic realizations of the dephasing dynamics of the linear  model~\eqref{eq:H_defect} with $\kappa=2$,  $\beta = 1$ so that $\gamma = 2/3$. Note that relaxation is the fastest for the red curve with $\epsilon=C$.}
\label{fig:rate_vs_eps}
\end{figure}

\subsection{Exact results for the trimer}

We now specialize to a trimer system with three-nodes ($N=3$) in the presence of a single on-site defect where some exact analytical results can be provided. 
Owing to lattice periodicity, the specific location of the defect is irrelevant.
The eigenvalues $E_\nu$ and eigenvectors $\ket{\xi^\nu}$ of the Hamiltonian $H$, obtained from Eqs.~\eqref{eq:H-eigval-first-set}--\eqref{eq:H-eigenvec-remain}, are substituted into Eq.~\eqref{eq:W-def} to construct the overlap matrix $W$ [see Appendix~\ref{sec:app-3-lev}].
Evaluating the eigenvalue $\lambda_2$ yields the relaxation time
\begin{align}
\tau_{\mathrm{relax}} = \frac{\epsilon^2 - 2 \epsilon C + 9 C^2}{6\gamma C^2} ,
\label{eq:3lev-tau}
\end{align}
which exhibits a nonmonotonic dependence on the defect strength $\epsilon$, attaining a minimum value at $4/(3\gamma)$ at  $\epsilon= C$.
This behavior is illustrated in Fig.~\ref{fig:rate_vs_eps}(a), where the scaled relaxation rate $\mu_2/\gamma = 1/(\gamma \, \tau_{\mathrm{relax}})$ reaches its maximum value $3/4$ at $\epsilon = C$.
Note that for small but non-zero defect strength $\epsilon$, one has $\tau_{\mathrm{relax}}\simeq 3/(2\gamma)$ and $\mu_2/\gamma \simeq 2/3$. 
Figure~\ref{fig:rate_vs_eps}(b) shows the time evolution of the scaled total energy for different values of $\epsilon$, clearly demonstrating that energy relaxation is fastest when $\epsilon = C$.

\begin{figure}[t]
\centering
\includegraphics[scale=0.91]{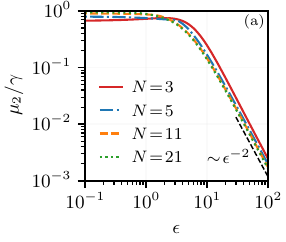}\hskip-2pt
\includegraphics[scale=0.91]{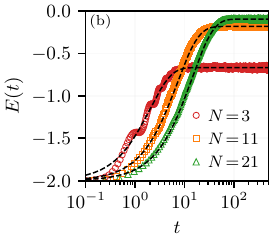}
\caption{(a) Scaled relaxation rate $\mu_2$ as a function of $\epsilon$ for different values of $N$. Lines are obtained by numerically computing the eigenvalue $\lambda_2$ of $W$ with $C=2$. (b) Evolution of total energy as a function of time $t$ for different values of $N$. The points are obtained from $2^{15}$ stochastic realizations of the dephasing dynamics of  model~\eqref{eq:H_defect} with $C=\epsilon=\kappa=2$ and $\beta = 1$. The dashed lines denote the analytical form $E(t) =  E(\infty) + (E(0) - E(\infty)) e^{-t/\tau_{\mathrm{relax}}}$.}
\label{fig:relaxation_linear}
\end{figure}

\subsection{Comparison with stochastic simulations}

To further assess the validity of Eq.~(\ref{eq:Idot}) for larger $N$, we compare it with direct simulation of the dynamics.   
For sufficiently large defect strength, the relaxation rate decreases algebraically.
As shown in Fig.~\ref{fig:relaxation_linear}(a), the scaled rate $\mu_2/\gamma$ decays as $\epsilon^{-2}$ for increasing $\epsilon$.
Remarkably, this trend is already captured analytically for the trimer system by Eq.~\eqref{eq:3lev-tau}, which yields $\tau_{\mathrm{relax}} \simeq {\epsilon^2}/{6 \gamma C^2}$ for large $\epsilon$.
Figure~\ref{fig:relaxation_linear}(a) also suggests that for small  defect strength, the relaxation rate is slightly reduced compared to the defect-free value $\mu_2=\gamma$, indicating a modest slowing down of relaxation. 
However, as $N$ increases, this effect becomes progressively weaker and the relaxation rate approaches its defect-free limit, reflecting the fact that the influence of a small local defect becomes negligible in large systems.
Figure~\ref{fig:relaxation_linear}(b) illustrates the time evolution of the total energy $E(t)$ for different system sizes $N$.
In all cases, the numerical data are well described by an exponential relaxation, $E(t)=E(\infty)+\big[E(0)-E(\infty)\big]e^{-t/\tau_{\mathrm{relax}}}$, which reflects incoherent energy exchange between eigenstates mediated by dephasing, with the slowest mode controlling the long-time dynamics.
The relaxation time $\tau_{\mathrm{relax}}$ extracted from the dynamics is set by the eigenvalue $\lambda_2$ of the matrix $W$, in full agreement with the analytical prediction.

\begin{figure}[t]
\centering
\includegraphics[scale=1]{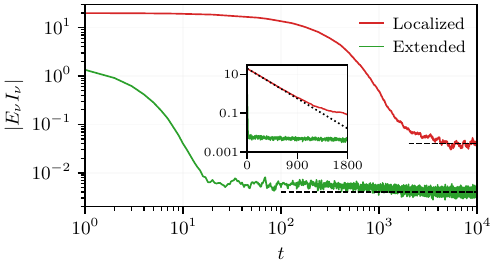}
\caption{Relaxation of the absolute value of mode energy $|E_\nu I_\nu|$ in a chain of $N=501$ sites with $C=1$ and $\epsilon=20$. Two types of initial conditions are compared: one in which all the initial energy is fed into the localized mode (red solid line) with $E_\nu \approx -20.09975$ and one where an extended mode  with $E_\nu \approx -1.99996$ is initially excited (green solid line). 
The dashed lines show the steady-state value of the mode energy, i.e., $|E_\nu|/ N$. 
The inset plot shows the same quantities in the semi-log scale. Trajectories are generated directly in mode space, using Eq.(\ref{eq:modedyn}) with uniform $g(\theta)$ so that $\kappa=2$ and an exponential $p(\tau)$ with $\beta=100$, and averaged over 100 stochastic realizations.
The black dotted line in the inset depicts
the expected exponential decay,  
$\exp(-\mu_2 t)$, computed as explained in the text.}
\label{fig:decay}
\end{figure}

To further test the above results, we consider a larger lattice and a relatively large defect strength $\epsilon$, and generate trajectories using Eq.~\eqref{eq:modedyn}, which allows us to monitor the dynamics directly in the action space.
We focus on initial conditions in which a single mode is excited, namely, 
$a_\nu(0)\propto\delta_{\nu\nu_0}$.
In particular, we compare two cases: one in which $\nu_0$ corresponds to an eigenstate exponentially localized at the defect site, and one in which $\nu_0$ corresponds to an extended mode (for example, see Fig.~\ref{fig:H-defect-spectrum}(b)).
As shown in Fig.~\ref{fig:decay}, these two choices lead to a dramatic separation of relaxation time scales, with the initially localized mode relaxing significantly more slowly than the extended one.
The observed decay is well described by an exponential law with a rate $\mu_2$ predicted by the kinetic approach (see inset of Fig.\ref{fig:decay}).

%================================================
%================================================
%================================================
\section{Large-deviation approach}
\label{sec:largeD}

To assess the dynamical relevance of trajectories leading to slow thermalization, we rely on the large-deviation theory. 
Over the past decades, this framework has emerged as a powerful tool for characterizing fluctuations in stochastic processes and nonequilibrium statistical mechanics~\cite{touchette_large_2009,touchette_introduction_2018,jack_ergodicity_2020}.  
In dynamical systems, large-deviation theory provides a means to compute the probability of atypical trajectories and the value of observables that depart significantly from their average, typical ones. 
Non-analyticities in the corresponding large-deviation functions signal \emph{dynamical phase transitions}, reflecting qualitative changes in the statistical properties of trajectories.
In the present context, this approach allows us to quantify and characterize regions in phase space that yield possibly slow, or at least substantially different from average, relaxation pathways, thereby offering insight on the different phase-space trajectories followed during equilibration. 

The main starting point of our analysis is the observation that Eq.~\eqref{eq:Idot} describes a continuous-time random walk of quasi-particles in action space. 
The corresponding transition rates are not assigned a priori but are fully determined by the eigenstate structure of the underlying linear problem through the matrix $W$.
This naturally defines an \textit{action network}, whose connectivity provides information on the out-of-equilibrium dynamics~\cite{lepri_thermalization_2023}.
Within the standard formalism developed for large-deviation analysis of Markov processes~\cite{touchette_large_2009}, the dephasing-induced dynamics~\eqref{eq:Idot} can be recast as a classical rate equation for the mode populations, 
\begin{align}
    \dot{\overline I}_\nu =\sum_{\mu} \Big( R_{\nu\mu}\,\overline I_\mu - R_{\mu\nu}\,\overline I_\nu \Big) ,
\label{eq:Idot_R}
\end{align}
where the transition-rate matrix is given by $R_{\nu\mu}=\gamma(1-\delta_{\nu\mu})\,W_{\nu\mu}$.
Such a representation makes explicit the stochastic jump structure of the dynamics and provides a natural starting point for applying large-deviation techniques to characterize rare trajectories and slow relaxation pathways. 

As a dynamical observable, let us consider the \emph{activity}, defined as the total number $K$ of jumps, corresponding to configuration changes between modes, occurring along a trajectory over an observation time~$t$.
To compute its large deviations, one may consider  
the tilted generator $W_K(s) $ defined by the $N\times N$ matrix with elements~\cite{lecomte_thermodynamic_2007,garrahan_dynamical_2007,garrahan_first-order_2009} 
\begin{equation}
\left( W_K(s) \right)_{{\mu}\nu}\equiv e^{-s}R_{{\mu} \nu}  - r_\nu \,\delta_{\nu \mu}, 
\end{equation}
with the escape rate from state $\nu$ given by $r_\nu \equiv \sum_{\mu \neq \nu}R_{{\mu} \nu} $.
Note that the tilted operator $W_K(s)$ has  $-r_\nu$ on the diagonal and the  first terms appears only in off-diagonal entries (see Ref.~\cite{lecomte_thermodynamic_2007} for details).  
The large-deviation  statistics is encoded in the leading eigenvalue and (right and left) eigenvectors of the tilted operator (also called scaled cumulant generating function). 
More precisely, we denote by $\lambda_K(s,N)$ the largest eigenvalue of $W_K(s)$ along with the corresponding right eigenvector $\phi_K$, defined by $W_K(s)\phi_K \equiv \lambda_K\phi_K$. Upon normalization, $\phi_K$ can be considered as a steady  probability distribution for the ``biased'' dynamics (sometimes  termed $s$--ensemble).
As it is known~\cite{lecomte_thermodynamic_2007}, for  $s = 0$ the eigenvector yields the density in the steady state, which in our case is the equipartition which is homogeneous. 
For $s < 0$, it probes the regime in which the mean activity $K /t$ of histories is typically larger than
in the steady state, while for $s > 0$ histories with smaller $K /t$ are favored. 
As an order parameter, we consider~\cite{lecomte_thermodynamic_2007}
\begin{equation}
\rho_K(s) = \frac{1}{N}\sum_{\nu=1}^N \nu \, {\phi_{K,\nu}}(s) ,
\label{ordpar}
\end{equation}
where $\phi_{K,\nu}(s)$ denotes the $\nu$th component of the eigenvector $\phi_K(s)$. 
The quantity $\rho_K(s)$ measures the occupancy of the steady state and thus of the distribution of modes that are involved in the biased dynamics.

\begin{figure}[t]
\centering
\includegraphics[scale=0.99]{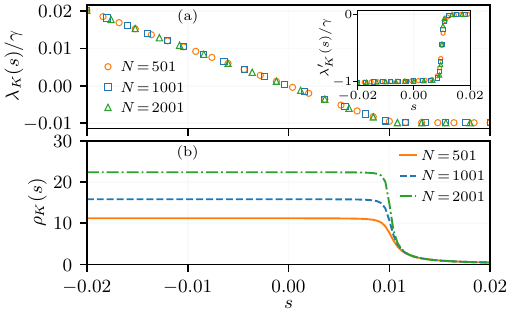}
\includegraphics[scale=1]{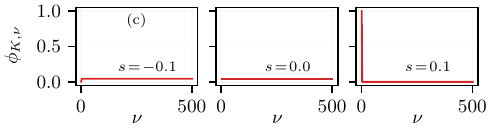}
\caption{Large deviations of activity for the defective Hamiltonian~\eqref{eq:H_defect} with $\epsilon=20$, $C=1$ and $\kappa=2$, $\beta=1$: 
(a) the rescaled largest eigenvalue $\lambda_K(s)$ of $W_K(s)$ as 
a function of $s$  for different chain lengths $N$; 
Inset: the numerical derivative $\lambda'_K(s)/\gamma$ 
computed by finite differences;
(b) the order parameter $\rho_K(s)$; (c) the components of eigenvectors 
$\phi_K$ for $N=501$ and three different values of $s$. Note the qualitative
changes from extended ($s=-0.1$ and $0$) to localized ($s=0.1$).}
\label{fig:activity_defect}
\end{figure}

In Fig.~\ref{fig:activity_defect}(a), we report the scaled large deviation function $\lambda_K(s)/\gamma$ for different system sizes, obtained by computing the leading eigenvector numerically for a relatively 
large $\epsilon$. 
The curves show good data-collapse.
Moreover, there is a sharp change in the slope around a characteristic value $s=s_*~(\simeq 0.01)$ where the derivative of $\lambda_K(s)$ changes abruptly (see the inset of Fig.~\ref{fig:activity_defect}(a)). 
Although there is not a real discontinuity, this signals the existence of two qualitative sets of trajectories: one with a finite activity, meaning a finite rate of energy exchange among normal modes, and another with little or very slow dynamics. 
This is fully consistent with our analysis showing that initial conditions localized around the defect site thermalize much slowly than extended ones, see Fig.~\ref{fig:decay}.

A complementary view is provided by the order parameter $\rho_K(s)$, which also displays a qualitative change, with a step discontinuity at $s=s_*$, see Fig.~\ref{fig:activity_defect}(b). 
In turn, the size dependence of $\rho_K(s)$  reveals a macroscopic contribution  of finite-activity degrees of freedom involved in biased dynamics, while the low-activity component remains finite upon increasing $N$. 

Moreover, the structure of the leading eigenvector $\phi_K$, changes from uniform to localized around $s_*$, see  Fig.~\ref{fig:activity_defect}(c).
An appealing way to understand such a qualitative change is to describe it as a form of condensation occurring in the action (or mode) space \cite{zannetti_condensation_2014,corberi_probability_2019}. In other words, most trajectories (a macroscopic fraction) in the localized phase concentrate on a subset of the action space. 
The shape of the eigenvector then reveals the actual modes mostly involved in the dynamics (the localized one in the present case).

\begin{figure}[t]
\centering
\includegraphics[scale=1.1]{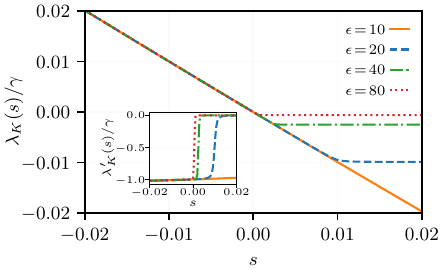}
\caption{Large deviations of activity  in terms of the rescaled largest eigenvalue $\lambda_K(s)$ of $W_K(s)$ as  a function of $s$ for increasing values of the parameter $\epsilon$. Here, we have $N=501$, $C=1$ for the defective Hamiltonian~\eqref{eq:H_defect} with $\kappa=2$ and  $\beta=1$. The inset plot shows  the numerical derivative $\lambda'_K(s)/\gamma$  computed by finite differences for the same  set of $\epsilon$ values.}
\label{fig:activity_epsilon}
\end{figure}

An interesting feature emerges in the large $\epsilon$ limit.
Since the separation of time-scales becomes increasingly large with increasing $\epsilon$, we expect that the large deviation function should reflect this. 
Indeed, as shown in Fig.~\ref{fig:activity_epsilon}, the position of the crossover value $s_*(\epsilon)$ gradually shifts towards zero upon increasing $\epsilon$. Moreover, the derivative $\lambda'_K(s)$ changes  more and more abruptly as $s_*$ approaches zero, as shown in the inset plot of Fig.~\ref{fig:activity_epsilon}.

Based on these numerical results, we may argue that this is compatible with a real jump discontinuity of $\lambda_K(s)$ in $s=0$ for infinite defect strength. 
In other words, there may exist a dynamical phase transition at $\epsilon=\infty$.
From a large-deviation perspective, analogous dynamical first-order phase transitions are often observed in several contexts, like for instance the dynamics of run-and-tumble particles~\cite{gradenigo_first-order_2019}, or in models of glasses~\cite{garrahan_dynamical_2007,jack_ergodicity_2020}. 
The usual interpretation for a singularity at $s=0$ is that there are two phases, an active one for $s < 0$ and an inactive one for $s > 0$.
Physical dynamics takes place at $s=0$, where the two dynamic phases coexist.  

%================================================
%================================================
%================================================
\section{Nonlinear defects}
\label{sec:nonlinDefects}

We now incorporate nonlinear interactions into the tight-binding chain and examine their impact on the relaxation dynamics.
We begin by considering a lattice with a single nonlinear defect, where the nonlinear cubic term is confined to one lattice site, while the remainder of the chain remains linear; see, e.g., Refs.~\cite{molina_nonlinear_1993,tsironis_generalized_1994,brazhnyi_spontaneous_2011,lepri_asymmetric_2011,dambroise_eigenstates_2013}. 
This construction provides a minimal extension of the linear defect model and allows us to isolate the impact of localized nonlinearity on the dynamics in the presence of dephasing noise. 
We then generalize the analysis to the fully nonlinear case, leading to the discrete nonlinear Schr\"odinger (DNLS) equation with interactions present on all sites. 
We finally discuss the role of higher-order nonlinearities for the relaxation process.

Unlike the linear defect model, where dephasing leads to a closed kinetic description in terms of the overlap matrix $W$ and a corresponding Markov process in action space, no analogous analytical framework is presently available for nonlinear defects. 
Consequently, our investigation here relies primarily on numerical simulations, supplemented by phenomenological arguments.
In addition, the tilted-generator analysis based on the large-deviation approach developed in Sec.~\ref{sec:largeD}, which allowed us to characterize rare relaxation pathways in the linear case, cannot be straightforwardly extended to the nonlinear problem.   
Rare and extreme events are nevertheless known to play an important role in nonlinear lattice dynamics.
Extreme events in discrete nonlinear lattices are often characterized by heavy-tailed distributions and extreme-value statistics, particularly in regimes close to integrability~\cite{maluckov_extreme_2009,maluckov_extreme_2013}. 
Explicit signatures of large deviation behavior have also been identified in the intermittent energy flux of DNLS chains, where the irregular birth and death of discrete breathers are associated with rare-event statistics~\cite{iubini_chain_2017}. 
Moreover, such extreme structures can be highly sensitive to perturbations, which may induce a crossover from transient rogue-wave events to persistent breathing patterns~\cite{hoffmann_extreme_2017}. 
While these studies focus on rare localized excitations, extreme events, and breather-mediated dynamics, our objective here is different. We investigate how the dephasing-induced relaxation mechanisms identified in the previous Sec.~\ref{sec:dephasing} for linear defects are modified when the defect becomes amplitude dependent, and we use the DNLS framework primarily as a nonlinear generalization of the defect problem.

\subsection{Single nonlinear defect}
\label{ssec:snd}

We consider the simplest nonlinear extension of Eq.~(\ref{eq:dls+def}) by replacing the defect term $\epsilon \psi_M$ at site $M$ with a cubic nonlinearity, namely
\begin{equation}
\label{eq:sd}
\ii \dv{\psi_j}{t} = -C  ( \psi_{j+1}+\psi_{j-1}  ) - \delta_{j M}\,|\psi_M|^2\psi_M\,.  
\end{equation}
In the following, we will refer to this model as the Single Nonlinear Defect (SND) model.
Upon recognizing that $\{\ii \psi_n^*$, $\psi_n\}$ form a proper set of (classical) canonical variables, Eq.~(\ref{eq:sd}) can be obtained from the Hamiltonian function
\begin{align}
\label{eq:h_snd}
H_{\mathrm{SND}} = - C \sum_j\left( \psi_j^* \psi_{j+1} + \psi_{j+1}^* \psi_j \right) - \frac{1}{2}|\psi_M|^4   
\end{align}
via the Hamilton equations $\dot\psi_j = \partial H / \partial (\ii \psi_j^*)$.
In addition to the Hamiltonian dynamics here considered, dephasing noise is implemented in the same way as in Sec.~\ref{ssec:kin}. 
As a result, the total norm defined as $A \equiv \sum_{j=1}^N |\psi_j|^2$ is exactly conserved, while the total energy $H$ is not.

We will focus on the case, where the defect site is initialized with a large local norm $|\psi_M(0)|^2=b\gg 1$, while all remaining lattice sites are empty. Under the action of the dephasing noise, energy and norm are progressively transferred from the defect site to the whole chain.

To establish a proper comparison with the full linear model with defect, we consider Eq.~(\ref{eq:dls+def}) with $\epsilon=b$ and initialized in the same kind of localized state, $|\psi_j(t=0)|^2=b \,\delta_{M 0} $. 
Accordingly, the frequencies of both localized and extended linear modes initially coincide in the two models. 
Note that the total norm $A=b$ here is not normalized: this choice naturally allows to introduce the perturbative parameter $b^{-1}$, see below.

\begin{figure}[t]
\centering
\includegraphics[scale=1]{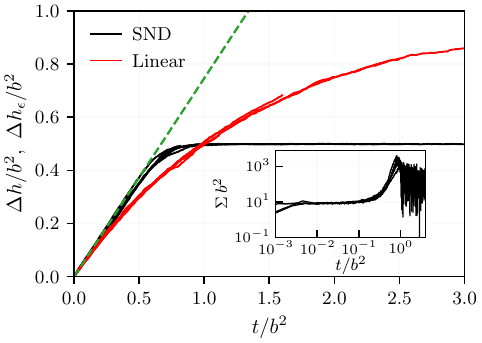}
\caption{Average energy variation $\Delta h(t)$ for SND model (black lines) and $\Delta h_\epsilon(t)$ for linear chain (red lines) for different values of the initial mass $b=\{15,20,25,30,35\}$. Dashed line refers to a linear fit of SND data. Averages are computed over a sample of 100 independent trajectories. Chain length $N=31$, $C=1$.  Noise parameters $\beta=1$ and $f=0.1$. Inset: evolution of the relative energy fluctuations for the SND model.}
\label{fig:comp2}
\end{figure}

In Fig.~\ref{fig:comp2} we show the time evolution of the (absolute) average  energy variation of the defect site $\Delta h(t)=1/2\langle \,|\psi_M(0)|^4-|\psi_M(t)|^4\,\rangle$
of the SND model (black lines) compared to the analogous energy variation in the linear model $\Delta h_\epsilon(t)= \langle h_\epsilon(0)-h_\epsilon(t)\rangle$
(red lines). 
Different values of the initial defect norms $b$ are considered and the axes are properly rescaled to obtain a data collapse.
In the linear model and for $b\gg 1$ the defect energy follows the relation $\Delta h_\epsilon(t)=h_\epsilon(0)(1-\exp(t/\tau_{relax}))$,  where 
$\tau_{relax}\simeq b^2$, in agreement with the results in Sec.~\ref{ssec:kin}. 
In the SND model, energy relaxation follows a linear relation $\Delta h=k t$, where $k$ is independent of $b$. 
Remarkably, the resulting relaxation is faster than that of the linear model, while it is well approximated in the short-time regime $t/b^2\ll 1$.

This result can be qualitatively rationalized by noting that for a generic time $t>0$ during the relaxation process, the nonlinear defect experiences a local norm $|\psi_M (t)|^2 < b$. As a result, the corresponding relaxation timescales are faster than those of the linear model constructed with a 
{\it fixed} $\epsilon$ given by the initial defect frequency $\epsilon=b$. More precisely, the passage from an exponential relaxation law for
$\Delta h_\epsilon(t)$ in the linear model to a linear law for $\Delta h(t)$ in the SND model can be justified as follows.

Within the linear system, from Sec.~\ref{ssec:kin}, the relevant relaxation dynamics of the defect energy $\langle h_\epsilon(t)\rangle $ is ruled by
\begin{equation}
\label{eq:dhl}
    \dv{\! \expval{h_\epsilon}}{t}= - \frac{\gamma}{\epsilon^2} \langle  h_\epsilon\rangle\,.
\end{equation}
In the spirit of an adiabatic approximation, we can now promote $\epsilon$ to be a function of $h_\epsilon(t)$ itself. 
Consistently, we impose $\epsilon=\langle |\psi_M(t)|^2 \rangle=\sqrt{\langle h_\epsilon(t) \rangle }$. 
The resulting equation $\mathrm{d} \langle h_\epsilon \rangle / \mathrm{d} t = -\gamma $ adequately accounts for a linear decay of $\langle h_\epsilon(t)\rangle$ and corresponds to the linear expansion of the solution of Eq.~(\ref{eq:dhl}) for $\gamma t\epsilon^{-2}\ll 1$.
As a further step, one can approximate $\langle h_\epsilon \rangle = \langle |\psi_M|^2 \rangle^2 \simeq \langle |\psi_M|^4\rangle $, assuming that
fluctuations of $|\psi_M|^2$ are negligible with respect to $\langle |\psi_M|^2 \rangle$. As a result, an analogous linear decay is expected for $\langle |\psi_M|^4(t)\rangle$ and therefore for $\Delta h (t)$. 
In the inset of Fig.~\ref{fig:comp2}, we estimate for the SND model the relative impact of energy 
fluctuations
\begin{equation}
\Sigma= \frac{ \langle |\psi_M|^4(t)\rangle - \langle |\psi_M|^2(t) \rangle^2} { \langle |\psi_M|^2(t) \rangle^2} \,.
\end{equation}
The data collapse indicates that $\Sigma$ scales as $b^{-2}$ in a wide range of rescaled times $t/b^2$. More precisely, we found that $\Sigma$ lies between $10^{-2}$ and $10^{-1}$ for $t/b^2 \lesssim 0.5$ for all values of $b$ considered here, which confirms our assumptions.

\subsection{Discrete Nonlinear Schr\"odinger Equation}

Upon incorporating local nonlinearities to all lattice sites in Eq.~(\ref{eq:sd}), we obtain the celebrated 
DNLS equation~\cite{kevrekidis_discrete_2009}
\begin{align}
\ii \dv{\psi_j}{t} = -C (\psi_{j+1} + \psi_{j-1}) -|\psi_j|^2 \psi_j\,.
\end{align}
Nonlinear defects in the form of large-amplitude peaks $|\psi_j|^2=b \gg 1$ can now be created non-preferentially on all
sites of the chain (with periodic boundary conditions).  They correspond to the so-called discrete breather solutions,  which have been
widely studied in the literature~\cite{flach_discrete_1998,flach_discrete_2008}. 
Other studies have also investigated the effect of explicit linear and nonlinear impurities on scattering or stability problems~\cite{kevrekidis_instabilities_2003,morales-molina_trapping_2006,palmero_solitons_2008,hennig_nonlinear_2025}.
Statistically, localized states have also been associated with a condensation phenomenon through large-deviation techniques~\cite{gradenigo_localization_2021}.
    
The problem of breather relaxation in the DNLS equation displays peculiar features. 
If the breather evolves either in an isolated chain or sufficiently far from external stochastic sources, then its relaxation timescales are found to be exponentially long in $b$ due to the existence of an adiabatic invariant of dynamical origin. 
On the other hand, when the breather is subject to direct stochastic noise, its evolution is much faster because the adiabatic invariant is broken~\cite{iubini_dynamical_2019,politi_frozen_2022}. 

The case of relaxation with uniform dephasing noise was recently studied in~\cite{ebrahimi_dynamics_2025}, where a linear relaxation law was found for the energy transfer mechanism.
An analogous linear scaling was also obtained in fully stochastic versions of the DNLS model~\cite{iubini_coarsening_2014,iubini_relaxation_2017}.
A different norm-preserving stochastic description representative of a finite-temperature DNLS evolution was discussed in~\cite{ebrahimi_stochastic_2025}.
Here we will focus our analysis on the case of uniform dephasing noise introduced in Sec.~\ref{sec:dephasing} and limit our study to the case of a single large-amplitude excitation located at site $j=M$.  

\begin{figure}[t]
\centering
\includegraphics[scale=1]{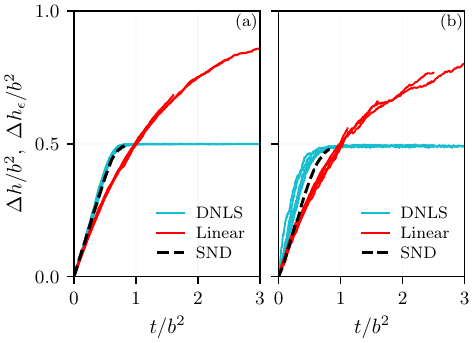}
\caption{Average energy variation $\Delta h(t)$ for the DNLS equation (solid cyan lines) and SND model (dashed black lines) compared with corresponding variations $\Delta h_\epsilon(t)$ of linear chain (solid red lines) for different values of the initial mass $b=\{15,20,25,30,35\}$ at site $j=M$. 
Panel (a) initially empty chain $\psi_{j\neq M}=0$; panel (b) Poissonian distribution of norms and random phases, see text.
Averages are computed over a sample of 100 independent trajectories. Chain length $N=31$, $C=1$.  Noise parameters $\beta=1$ and $f=0.1$.}
\label{fig:comp_dnls}
\end{figure}

In the same spirit as in Fig.~\ref{fig:comp2},
Fig.~\ref{fig:comp_dnls} shows the evolution of the peak's average energy variation $\Delta h$ in the DNLS equation (solid cyan lines) compared with the 
corresponding tight-binding model with defect (solid red lines), for different values of the peak norm $b$ at site $j=M$. 
For further comparison, we show the SND dynamics as dashed black curves.
For all cases, two distinct initial conditions for 
the remaining lattice variables $\psi_j\,, j\neq M$ are considered: in panel (a) they are all vanishing (empty chain); in panel (b) they are chosen
according to a Poissonian distribution of norms (with unit average), $P(|\psi_j|^2) \sim \exp(-|\psi_j|^2)$, and uniformly distributed phases in the interval $[0,2\pi)$.
The latter state in the DNLS model 
corresponds to an equilibrium state at infinite temperature~\cite{rasmussen_statistical_2000,rumpf_simple_2004,ebrahimi_dynamics_2025}.
We find that DNLS and SND dynamics are practically indistinguishable for the initially empty chain  and, in turn, well described by the linear model for short times. 
On the other hand, for the infinite-temperature initial conditions, a slight speed-up of the DNLS relaxation process is found with respect to the other two models.
We interpret this result as a manifestation of intrinsically nonlinear energy-transfer processes occurring in the DNLS model~\cite{kenkre_self-trapping_1986,rumpf_intermittent_2004} that cannot be properly described in terms of the dynamics of linear chains with linear or nonlinear defects. 

\subsection{Generalized nonlinearities}
We conclude this section with some remarks on DNLS lattices with higher-order nonlinearities~\cite{johansson_statistical_2004}
\begin{align}
\label{eq:dnls_alpha}
\ii \dv{\psi_j}{t} = -C (\psi_{j+1} + \psi_{j-1}) -|\psi_j|^{2(\alpha-1)} \psi_j
\end{align}
with $\alpha>1$~\footnote{The case $\alpha<1$ implies that the absolute defect energy increases slower than its norm $b$ for large $b$, a regime, which is beyond the scope of the present study.}. 
We consider again the relaxation dynamics corresponding to initializing the lattice with a single peak with norm $|\psi_M|^2=b$ on an empty chain.
Generalizing the arguments of Sec.~\ref{ssec:snd}, we identify the closest linear tight-binding model with defect upon choosing 
$\epsilon=|\psi_M(0)|^{2(\alpha-1)}$.
We can now revisit Eq.~(\ref{eq:dhl}) for the evolution of the defect energy with the adiabatic condition $\epsilon=\langle h_\epsilon\rangle ^{1-1/\alpha}$,
which is equivalent to $\langle h_\epsilon \rangle = \langle |\psi_M|^2 \rangle ^\alpha$.
Accordingly, one obtains the equation
\begin{equation}
     \frac{\mathrm{d}\langle h_\epsilon \rangle}{\mathrm{d}t} = -\gamma  \langle h_\epsilon \rangle^{\left(\frac{2}{\alpha}-1\right)}\,
\end{equation}
whose solution is
\begin{equation}
\label{eq:dh_alpha}
     \langle h_\epsilon\rangle ^{\left(2-\frac{2}{\alpha}\right)} (0) - \langle h_\epsilon \rangle^{\left(2-\frac{2}{\alpha}\right)}(t) = \frac{\gamma t}{2-\frac{2}{\alpha}}\,.
\end{equation}
Equation~(\ref{eq:dh_alpha}) clarifies that for generalized DNLS models the defect energy $\langle h_\epsilon(t) \rangle$ evolves nonlinearly in time. Only for the 
standard cubic DNLS $(\alpha=2)$, $\langle h_\epsilon(t) \rangle$ is a linear function of time. Remarkably, in the limit $\alpha \to +\infty$, Eq.~(\ref{eq:dh_alpha})
predicts a square-root scaling of $\langle h_\epsilon(t)\rangle$ with $t$.
Finally, assuming $\langle |\psi_M|^{2\alpha}\rangle \simeq \langle |\psi_M|^2 \rangle^\alpha$, we can extend Eq.~(\ref{eq:dh_alpha}) to the evolution of  $\langle |\psi_M|^{2\alpha}\rangle$ in place of $\langle h_\epsilon\rangle$.

\begin{figure}[t]
\centering
\includegraphics[scale=1]{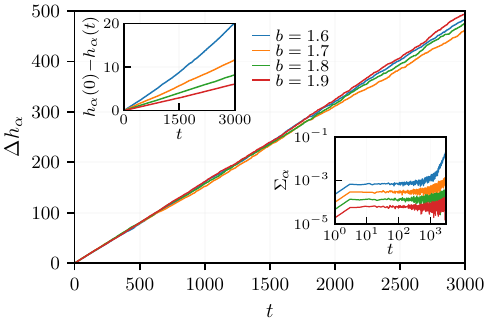}
\caption{Average variation $\Delta h_\alpha(t)$ for a generalized DNLS equation, Eq.~(\ref{eq:dnls_alpha}), with $\alpha=8$ and for different values of the initial defect norm $b$.
Upper inset: behavior of defect energy difference for the same data. 
Lower inset: evolution of generalized relative fluctuations $\Sigma_\alpha$, see text.  Averages are computed over a sample of 100 independent trajectories.  
Chain length $N=31$, $C=1$.  Noise parameters $\beta=1$ and $f=0.1$.}
\label{fig:gen_dnls}
\end{figure}

In Fig.~\ref{fig:gen_dnls},  we compare the relaxation dynamics of a generalized DNLS model obtained numerically from Eq.~(\ref{eq:dnls_alpha}) with the estimate in Eq.~(\ref{eq:dh_alpha}).
We compute the quantity $\Delta h_\alpha = h_\alpha^{\left(2-\frac{2}{\alpha}\right)} (0) - h_\alpha^{\left(2-\frac{2}{\alpha}\right)}(t)$, where $h_\alpha(t)=\alpha^{-1}\langle |\psi_M|^{2\alpha} \rangle$ and specialize the analysis for $\alpha=8$. 
The evolution of $\Delta h_\alpha$ displays a rather clean linear behavior, with a slope independent of the initial norm $|\psi_M(0)|^2=b$, in agreement with Eq.~(\ref{eq:dh_alpha}). 
Conversely, a naive estimation of energy differences $h_\alpha(0)-h_\alpha(t)$ results in nonlinear $b$--dependent profiles; see the upper inset.
Finally, in the lower inset, we show the evolution of the generalized relative energy fluctuations
\begin{equation}
\Sigma_\alpha = \frac{ \langle |\psi_M|^{2\alpha}(t)\rangle - \langle |\psi_M|^2(t) \rangle^\alpha} { \langle |\psi_M|^2(t) \rangle^\alpha} \,.
\end{equation}
For the chosen parameters, $\Sigma_\alpha$ is always below $10^{-2}$, which confirms the validity of Eq.~(\ref{eq:dh_alpha}).

%================================================
%================================================
%================================================
\section{Conclusions}
\label{sec:conclu}

In this work, we studied the relaxation dynamics of a tight-binding chain subject
to local dephasing noise in the presence of defects.
We considered both linear and nonlinear defects, with analytical emphasis on the
case of a single linear on-site defect, for which an exact treatment is possible,
while nonlinear defects were investigated numerically.

For the linear defect, we derived closed-form expressions for the eigenvalues and
eigenvectors of the defective Hamiltonian~\eqref{eq:H_defect} in terms of those of
the defect-free system.
This enabled a formulation of the dephasing-induced dynamics entirely in the
eigenstate basis and identified the overlap matrix $W$ as the central object
governing relaxation.
Physically, the doubly stochastic matrix $W$ encodes the spatial overlap between
eigenstates and thus controls the efficiency of dephasing-induced intensity transfer.

We demonstrated that the stochastic kinetic approach based on random phase kicks is 
equivalent to a standard Lindblad description of local dephasing.
The kinetic formulation provides additional physical insight by explicitly relating
the effective dephasing rate to three independent sources of stochasticity:
waiting-time statistics, phase-kick distributions, and the random selection of
dephasing sites.
Both approaches lead to the same master equation for the mode populations, which
takes the form of a continuous-time random walk in action space.

Solving this master equation, we showed that relaxation toward the steady state is
generically exponential, with a rate set by the spectral gap of $W$.
Defects induce localized eigenstates with weak spatial overlap with extended modes,
creating bottlenecks for dephasing-induced transport in action space.
Consequently, relaxation slows down with increasing defect strength, and for strong
defects the relaxation rate scales as $\epsilon^{-2}$.
This behavior was confirmed analytically in small systems and numerically in larger
lattices, where a clear separation of relaxation time scales emerges depending on
whether localized or extended modes are initially excited.

To characterize relaxation pathways beyond typical behavior, we employed a
large-deviation approach to the stochastic dynamics in action space.
Within this framework, the dephasing-induced evolution is described as a random walk
on an \emph{action network} whose connectivity is fixed by the matrix $W$.
Using a tilted generator, we showed that trajectories separate into distinct
dynamical classes.
Low-activity trajectories are dominated by defect-localized modes and exhibit slow
energy redistribution, whereas high-activity trajectories involve extended modes
and relax more rapidly.
Increasing the defect strength enhances the separation between these regimes,
providing a unified interpretation of slow relaxation as arising from the interplay
between mode localization and rare fluctuations in action-space dynamics.

Finally, we extended our analysis to the case of nonlinear defects.
As a minimal nonlinear extension, we introduced a single nonlinear defect, for which
the defect energy variation is quantified by
$\Delta h(t)=\tfrac12\langle |\psi_M(0)|^4-|\psi_M(t)|^4\rangle$.
We compared this dynamics with that of the linear defect model, whose defect energy is
$h_\epsilon(t)=\epsilon|\psi_M(t)|^2$, 
choosing $\epsilon$ to match the initial local
frequency scales.
We found that energy relaxation at the nonlinear defect is quite well approximated by the
corresponding linear model at early relaxation stages, while it becomes significantly
faster for longer times.
This behavior can be understood by noting that, during relaxation, the local norm
$|\psi_M(t)|^2$ decreases in time, so that the nonlinear defect effectively
experiences a progressively weaker local energy scale, in contrast to the linear
case where $\epsilon$ remains fixed.
As a result, the exponential relaxation characteristic of linear defects crosses
over to an approximately linear-in-time decay of $\Delta h(t)$ over the dominant
relaxation window.
For the fully nonlinear DNLS dynamics, we observed relaxation behavior that closely
resembles the SND model for localized initial conditions, while thermal-like initial
states exhibit a mild additional speed-up, consistent with the presence of
intrinsically nonlinear energy-transfer processes.
This approach proved to be applicable also to the case of DNLS lattices with generalized nonlinearities.

Overall, our study provides a unified and physically transparent picture of
dephasing-induced relaxation in defective tight-binding systems, highlighting the
central role of eigenstate localization, action-space connectivity, and trajectory
fluctuations in shaping equilibration dynamics under noise.

\acknowledgments
We thank Stefano Ruffo for valuable discussions.
We acknowledge support from the MUR PRIN2022 project ``Breakdown of ergodicity in classical and quantum many-body systems'' (BECQuMB) Grant No. 20222BHC9Z.\\

\appendix

%================================================
%================================================
%================================================
\section{Spectrum of the defective tight-binding Hamiltonian}
\label{sec:app-defect-free-H}

In this appendix, we derive the eigenvalues $E_\nu$ and eigenvectors $\ket{\xi^\nu }$ of the defective tight-binding Hamiltonian $H$ given in Eq.~\eqref{eq:H_defect}.
To this end, we start with the defect-free tight-binding Hamiltonian
\begin{align}
    {H}_{\mathrm{TBC}} = -C \sum_{j=0}^{N-1} ( \ket{j}\bra{j+1} + \ket{j+1}\bra{j} ), 
\end{align}
with $C>0$ denoting the hopping amplitude and $\ket{N} = \ket{0}$. 
Solving the equation $H_{\mathrm{TBC}} \ket{\chi^\nu } = \omega_\nu \ket{\chi^\nu }$, one obtains 
the eigenvalues $\omega_\nu$ and the corresponding orthonormal eigenvectors $\ket{\chi^\nu }$, which are given in Eq.~\eqref{eq:specrum-freeH} of the main text.
As seen from Table~\ref{tab:tbm}, for odd $N$, there exists $(N-1)/2$ number of degenerate subspaces, which are denoted by 
$\mathcal D=\{\ket{\chi^\nu },\ket*{\chi^{N-\nu}} \}$.
For each of these degenerate subspaces, one may define a normalized vector
\begin{align}
\ket{\Sigma^\nu} &= \frac{1}{1+|b_\nu|^2} \qty( \ket{\chi^\nu }+ b_\nu \ket*{\chi^{N-\nu}} ) ,
\end{align}
where $b_\nu$ is a complex number, such that  $\ket{\Sigma^\nu}$ remains an eigenvector of the defect-free $H_{\mathrm{TBC}}$ with eigenvalue $\omega_\nu$. Moreover, for $\Sigma^\nu_M \equiv \braket{M}{\Sigma^\nu} = 0$, one has
$H \ket{\Sigma^\nu } = \omega_\nu \ket{\Sigma^\nu }$, i.e., $\ket{\Sigma^\nu }$   acts as eigenvector of the defective $H$ with eigenvalue $\omega_\nu$.
The condition $\Sigma^\nu_M = 0$ may always be satisfied by choosing $b_\nu = - \chi^{\nu}_M / \chi^{N-\nu}_M = - \chi^\nu_M / {(\chi^{\nu}_M)^*}$ with `$*$' denoting complex conjugation.
Consequently, for the Hamiltonian $H$, we may readily write $(N-1)/2$ eigenvalues and the corresponding eigenvectors, which are given in Eq.~\eqref{eq:H-eigval-first-set} of the main text.

For the remaining $(N+1)/2$ number of  eigenvectors $ \ket{\xi^\mu} $ with $\mu \in [0, (N+1)/2, (N+3)/2 ,\ldots (N-1)]$, we expand them in the defect-free eigenbasis as
\begin{align}
    \ket{\xi^\mu} = \sum_{\nu=0}^{N-1} c^{(\mu)}_\nu \ket{\chi^\nu } ,
    \label{eq:xi-expand-in-chi}
\end{align}  
where $c^{(\mu)}_\nu$ are complex scalars. Operating $ H $ on the both sides of Eq.~\eqref{eq:xi-expand-in-chi} gives
\begin{align}
    E_\mu  \ket{\xi^\mu}  &= \sum_\nu c^{(\mu)}_\nu  \omega_\nu \ket{\chi^\nu } - \epsilon S^{(\mu)} \ket{M} ,
    \label{eq:EmuXimu0}
\end{align}
where the scalar $S^{(\mu)}$ is defined by  $S^{(\mu)} \equiv \sum_\nu c^{(\mu)}_\nu \chi^{\nu}_M$.
Operating $ ({\chi}^{(\lambda)})^\dagger $ on Eq.~\eqref{eq:EmuXimu0} from the left and using the orthonormality of ${\chi}^{(\lambda)}$ yields
\begin{align}
c^{(\mu)}_\lambda = \frac{ \epsilon (\chi^{\lambda}_M)^* S^{(\mu)} }{ (\omega_\lambda - E_\mu )} ,
\label{eq:cmu-def}
\end{align}
which when substituted in the definition of $S^{(\mu)}$ gives a self-consistent equation
\begin{align}
S^{(\mu)} &=  S^{(\mu)} \frac{\epsilon}{N} \sum_{\lambda=0}^{N-1}  \frac{ 1 }{\omega_\lambda - E_\mu}  ,
\end{align}
where we have used $|\chi^{(\nu)}_M|^2 = 1/N$.

Assuming $S^{(\mu)}\neq 0$, we may then write a secular equation
\begin{align}
\sum_{\nu=0}^{N-1} \frac{1}{E_\mu - \omega_\nu } + \frac{N}{\epsilon} =0 ,
\label{eq:secular}
\end{align}
which, because of the existence of the $(N-1)/2$ doubly--degenerate eigenstates, reduces to an $(N+1)/2$--degree secular equation~\eqref{eq:secular1} given in the main text.
The roots of Eq.~\eqref{eq:secular1} yield the remaining $(N+1)/2$ eigenvalues $ E_\mu $ of the defective Hamiltonian $ H $.
For numerical purpose of root finding, it is convenient to multiply both sides of the secular equation~\eqref{eq:secular1} by the $((N+1)/2)$--degree polynomial $P(E_\mu) = \Pi_{\lambda=0}^{(N-1)/2} (E_\mu - \omega_\lambda )$ and rewrite it as
\begin{align}
   F(E_\mu) \equiv &\frac{N}{\epsilon} P(E_\mu) + \prod_{\lambda=1}^{(N-1)/2}  (E_\mu  - \omega_\lambda ) \nonumber \\
   &+ 2 \sum_{\nu=1}^{(N-1)/2}
        \prod_{\substack{\lambda=0 \\ \lambda\neq\nu}}^{(N-1)/2}  (E_\mu - \omega_\lambda) = 0.
\end{align}
Note that $F(E_\mu)$ is an $((N+1)/2)$--degree polynomial whose real roots gives the remaining  eigenvalues $ E_\mu $, and  the corresponding eigenvectors $ \ket{\xi^\mu} $, up to some normalization, may be written from Eqs.~\eqref{eq:xi-expand-in-chi} and~\eqref{eq:cmu-def} as
\begin{align}
\ket{\xi^\mu} =  \sum_{\lambda=0}^{N-1} \frac{(\chi^{\lambda}_M)^* }{\omega_\lambda - E_\mu} \ket*{\chi^\lambda} .
\end{align}
Using $|\chi_M^{(\lambda)}|^{2} = 1/N$, one writes
\begin{align}
  \braket*{ \xi^\nu  }{ \xi^\mu } = \frac{1}{N} \sum_{\lambda=0}^{N-1} \frac{1}
         {(\omega_\lambda-E_\mu)(\omega_\lambda-E_\nu)} ,
\end{align}
which  vanishes for $\mu\neq \nu$ due to the secular equation~\eqref{eq:secular} and  yields for $\nu=\mu$ the normalization of $\ket{\xi^\mu}$, such that we have 
$ \braket*{ \xi^\nu  }{ \xi^\mu } =  {\mathcal N}_\mu^{2} \delta_{\mu\nu}$
with
\begin{align}
   {\mathcal N}_\mu \equiv \displaystyle \qty[ \frac{1}{N} \sum_{\lambda=0}^{N-1} \frac{1} {(\omega_\lambda-E_\mu)^{2}}]^{1/2}  .
\end{align}
Hence, dividing ${{\xi}}^{(\mu)}$ by \(\mathcal{N}_\mu\) yields the remaining normalized eigenvectors, which are given in Eq.~\eqref{eq:H-eigenvec-remain} of the main text.

%================================================
%================================================
%================================================
\section{Relaxation in the trimer system with linear defect}
\label{sec:app-3-lev}

Let us consider a three-site $(N=3)$ defective tight-binding Hamiltonian with a defect at site $n=1$:
\begin{equation}
\label{eq:H_defect_3}
    {H} = -C \sum_{j=0}^{2} ( \ket{j}\bra{j+1} + \ket{j+1}\bra{j} ) - \epsilon \ket{1} \bra{1},
\end{equation}
The eigenvalues and eigenvectors for the defect-free case ($\epsilon =0$) obtained from Eq.~\eqref{eq:specrum-freeH} may be written as 
\begin{eqnarray}
&\omega_0=-2 C,  \quad & \ket*{\chi^{0}} = \frac{1}{\sqrt{3}} (1 , 1 ,  1 )^\top , \\
&\omega_1= C,  \quad &\ket*{\chi^{1}} = \frac{1}{\sqrt{3}} (1 , z ,  z^* )^\top , \\
&\omega_2= C,  \quad &\ket*{\chi^{2}} = \frac{1}{\sqrt{3}} (1 ,  z^*, z )^\top ,
\end{eqnarray}
where $z=e^{i2\pi/3}$.
Hence, for $\epsilon=0$, the only degenerate subspace is given by
$ {\mathcal D}=\{ \ket*{\chi^{1}}, \ket*{\chi^{2}} \} $. 
In the presence of the defect ($\epsilon \neq 0$), the eigenvector of $H$ in Eq.~\eqref{eq:H_defect_3} corresponding to the eigenvalue $E_1 = C$ obtained from Eq.~\eqref{eq:H-eigval-first-set} reads
\begin{align}
\label{eq:3levxi1}
    \ket*{ \xi^1} &= \frac{1}{\sqrt{2}}[\ket*{\chi^1} - (z/z^*) \ket*{\chi^2}]  
                  =\dfrac{1}{2\sqrt{2}} \left(
\begin{array}{c}
 \sqrt{3}+i \\
 0 \\
 -\sqrt{3}-i \\
\end{array}
\right).
\end{align}
The remaining two eigenvalues obtained from the secular equation~\eqref{eq:secular1}, which reduces to a quadratic equation for $N=3$, are given by
\begin{align}
    E_0 &= \frac{1}{2} \left(- C -\epsilon -\sqrt{9 C^2-2  \epsilon C +\epsilon ^2} \right) , \\ 
    E_2 &=  \frac{1}{2} \left(- C -\epsilon + \sqrt{9 C^2 - 2  \epsilon C +\epsilon ^2} \right) , 
\end{align}
while the corresponding normalized eigenvectors obtained from Eq.~\eqref{eq:H-eigenvec-remain} read
\begin{align}
\label{eq:3levxi0}
\ket*{\xi^0} &= \frac{1}{\epsilon \, \mathcal{N}_0} \left(
\begin{array}{c}
 \dfrac{2 C}{ \sqrt{9 C^2-2  \epsilon C +\epsilon ^2}-C +\epsilon } \\[0.5ex]
 1 \\[0.5ex]
 \dfrac{2 C}{ \sqrt{9 C^2-2  \epsilon C +\epsilon ^2}-C +\epsilon } \\
\end{array}
\right)
,  \\[1ex] 
\ket*{\xi^2} &= \frac{1}{\epsilon \,  \mathcal{N}_2} \left(
\begin{array}{c}
 \dfrac{ - 2 C}{ \sqrt{9 C^2-2  \epsilon C +\epsilon ^2}+ C -\epsilon } \\[0.5ex]
 1 \\[0.5ex]
 \dfrac{ - 2 C}{ \sqrt{9 C^2-2  \epsilon C +\epsilon ^2}+ C -\epsilon } \\
\end{array}
\right) ,
\end{align}
where
\begin{widetext}
\begin{align}
    \mathcal{N}_0 &= \qty[  \frac{4}{3} \left(\frac{2}{\left(\sqrt{9 C^2-2  \epsilon C +\epsilon ^2}+3 C +\epsilon \right)^2}+\frac{1}{\left(\sqrt{9 C^2-2  \epsilon C +\epsilon ^2}-3 C +\epsilon \right)^2}\right) ]^{1/2} , \\
    \mathcal{N}_2 &= \qty[ \frac{4}{3} \left(\frac{1}{\left(\sqrt{9 C^2-2  \epsilon C +\epsilon ^2}+3 C -\epsilon \right)^2}+\frac{2}{\left(\sqrt{9 C^2-2  \epsilon C  +\epsilon ^2} - 3 C -\epsilon \right)^2}\right)  ]^{1/2} .
\end{align}
The matrix $W$ for the three-node system~\eqref{eq:H_defect_3} under dephasing dynamics is obtained using Eq.~\eqref{eq:W-def} with Eqs.~\eqref{eq:3levxi0} and~\eqref{eq:3levxi1} and is given by
\begin{align}
    W = 
\begin{pmatrix}
 \frac{32 C^4+\left(\sqrt{9 C^2-2  \epsilon C +\epsilon ^2}-C +\epsilon \right)^4}{4 \left(9 C^2-2  \epsilon C +\epsilon ^2\right) \left(\sqrt{9 C^2-2  \epsilon C +\epsilon ^2}-C +\epsilon \right)^2} & \frac{\sqrt{9 C^2-2  \epsilon C +\epsilon ^2}+C-\epsilon }{4 \sqrt{9 C^2-2  \epsilon C +\epsilon ^2}} & \frac{3 C^2}{9 C^2-2  \epsilon C +\epsilon ^2} \\[2ex]
 \frac{\sqrt{9 C^2-2 \epsilon C +\epsilon ^2}+C-\epsilon }{4 \sqrt{9 C^2-2  \epsilon C +\epsilon ^2}} & \frac{1}{2} & \frac{\sqrt{9 C^2-2  \epsilon C +\epsilon ^2}-C+\epsilon }{4 \sqrt{9 C^2-2  \epsilon C +\epsilon ^2}} \\[2ex]
 \frac{3 C^2}{9 C^2-2 \epsilon C +\epsilon ^2} & \frac{\sqrt{9 C^2-2 \epsilon C +\epsilon ^2}-C+\epsilon }{4 \sqrt{9 C^2-2  \epsilon C +\epsilon ^2}} & \frac{32 C^4+\left(\sqrt{9 C^2-2  \epsilon C +\epsilon ^2}+C-\epsilon \right)^4}{4 \left(9 C^2-2 \epsilon C +\epsilon ^2\right) \left(\sqrt{9 C^2-2 \epsilon C +\epsilon ^2}+ C -\epsilon \right)^2} 
\end{pmatrix} .
\end{align}
The eigenvalues of $W$ are $\{\lambda_1, \lambda_2, \lambda_3 \}= \{1,1-\frac{6 C^2}{9 C^2-2  \epsilon C +\epsilon ^2},0 \}$, which gives $1-\lambda_2 \simeq 2/3$ for small nonzero values of $\epsilon$ and $1-\lambda_2 \simeq 6 C^2 / \epsilon^2$ for $\epsilon \gg C$.
The relaxation time obtained by $\tau_{\mathrm{relax}}= \gamma^{-1} (1-\lambda_2)^{-1}$ is given in Eq.~\eqref{eq:3lev-tau} of the main text.

\end{widetext}

%\bibliographystyle{unsrt}
%\bibliography{library}

\begin{thebibliography}{10}

\bibitem{ashcroft_solid_1976}
N.~W. Ashcroft and N.~D. Mermin.
\newblock {\em Solid {State} {Physics}}.
\newblock Holt, Rinehart and Winston, New York, 1976.

\bibitem{grosso_solid_2000}
G.~Grosso and G.~P. Parravicini.
\newblock {\em Solid {State} {Physics}}.
\newblock Academic Press, London, 2000.

\bibitem{christodoulides_discretizing_2003}
D.~N. Christodoulides, F.~Lederer, and Y.~Silberberg.
\newblock Discretizing light behaviour in linear and nonlinear waveguide
  lattices.
\newblock {\em Nature}, 424(6950):817--823, 2003.

\bibitem{morsch_dynamics_2006}
O.~Morsch and M.~Oberthaler.
\newblock Dynamics of {Bose}-{Einstein} condensates in optical lattices.
\newblock {\em Rev. Mod. Phys.}, 78(1):179--215, 2006.

\bibitem{lee_disordered_1985}
P.~A. Lee and T.~V. Ramakrishnan.
\newblock Disordered electronic systems.
\newblock {\em Rev. Mod. Phys.}, 57(2):287--337, 1985.

\bibitem{abrahams_50_2010}
E.~Abrahams.
\newblock {\em 50 {Years} of {Anderson} {Localization}}.
\newblock World Scientific, Singapore, 2010.

\bibitem{anderson_absence_1958}
P.~W. Anderson.
\newblock Absence of diffusion in certain random lattices.
\newblock {\em Phys. Rev.}, 109(5):1492--1505, 1958.

\bibitem{roati2008anderson}
G.~Roati, C.~D’Errico, L.~Fallani, M.~Fattori, C.~Fort, M.~Zaccanti,
  G.~Modugno, M.~Modugno, and M.~Inguscio.
\newblock Anderson localization of a non-interacting bose--einstein condensate.
\newblock {\em Nature}, 453(7197):895--898, 2008.

\bibitem{anderson_localized_1961}
P.~W. Anderson.
\newblock Localized magnetic states in metals.
\newblock {\em Phys. Rev.}, 124(1):41--53, 1961.

\bibitem{mott_theory_1961}
N.~F. Mott and W.~D. Twose.
\newblock The theory of impurity conduction.
\newblock {\em Adv. Phys.}, 10(38):107--163, 1961.

\bibitem{economou_greens_2006}
E.~N. Economou.
\newblock {\em Green’s {Functions} in {Quantum} {Physics}}, volume~7.
\newblock Springer, Berlin, 2006.

\bibitem{joannopoulos_photonic_2011}
J.~D. Joannopoulos, S.~G. Johnson, J.~N. Winn, and R.~D. Meade.
\newblock {\em Photonic {Crystals}: {Molding} the {Flow} of {Light} - {Second}
  {Edition}}.
\newblock Princeton University Press, New Jersey, 2011.

\bibitem{bruderer_probing_2006}
M.~Bruderer and D.~Jaksch.
\newblock Probing {BEC} phase fluctuations with atomic quantum dots.
\newblock {\em New J. Phys.}, 8(6):87, 2006.

\bibitem{klein_dynamics_2007}
A.~Klein, M.~Bruderer, S.~R. Clark, and D.~Jaksch.
\newblock Dynamics, dephasing and clustering of impurity atoms in
  {Bose}–{Einstein} condensates.
\newblock {\em New J. Phys.}, 9(11):411--411, 2007.

\bibitem{palzer_quantum_2009}
S.~Palzer, C.~Zipkes, C.~Sias, and M.~Köhl.
\newblock Quantum transport through a {Tonks}-{Girardeau} gas.
\newblock {\em Phys. Rev. Lett.}, 103(15):150601, 2009.

\bibitem{john_strong_1987}
S.~John.
\newblock Strong localization of photons in certain disordered dielectric
  superlattices.
\newblock {\em Phys. Rev. Lett.}, 58(23):2486--2489, 1987.

\bibitem{hewson_kondo_1993}
A.~C. Hewson.
\newblock {\em The {Kondo} {Problem} to {Heavy} {Fermions}}.
\newblock Cambridge University Press, Cambridge, 1993.

\bibitem{lepri_thermalization_2023}
S.~Lepri.
\newblock Thermalization of isolated harmonic networks under conservative
  noise.
\newblock {\em J. Stat. Phys.}, 190(1):16, 2023.

\bibitem{breuer_theory_2007}
H.~P. Breuer and F.~Petruccione.
\newblock {\em The {Theory} of {Open} {Quantum} {Systems}}.
\newblock Oxford University Press, 2007.

\bibitem{zurek_decoherence_2003}
W.~H Zurek.
\newblock Decoherence, einselection, and the quantum origins of the classical.
\newblock {\em Rev. Mod. Phys.}, 75(3):715--775, 2003.

\bibitem{lindblad_generators_1976}
G.~Lindblad.
\newblock On the generators of quantum dynamical semigroups.
\newblock {\em Commun. Math. Phys.}, 48(2):119--130, 1976.

\bibitem{gorini_completely_1976}
V.~Gorini, A.~Kossakowski, and E.~C.~G. Sudarshan.
\newblock Completely positive dynamical semigroups of \textit{{N}} -level
  systems.
\newblock {\em J. Math. Phys.}, 17(5):821--825, 1976.

\bibitem{chruscinski_brief_2017}
D.~Chruściński and S.~Pascazio.
\newblock A brief history of the {GKLS} equation.
\newblock {\em Open Syst. Inf. Dyn.}, 24(03):1740001, 2017.

\bibitem{manzano_short_2020}
D.~Manzano.
\newblock A short introduction to the {Lindblad} master equation.
\newblock {\em AIP Adv.}, 10(2):025106, 2020.

\bibitem{das_quantum_2022}
D.~Das and S.~Gupta.
\newblock Quantum random walk and tight-binding model subject to projective
  measurements at random times.
\newblock {\em J. Stat. Mech.: Theory Exp.}, 2022(3):033212, 2022.

\bibitem{das_quantum_2022-1}
D.~Das, S.~Dattagupta, and S.~Gupta.
\newblock Quantum unitary evolution interspersed with repeated non-unitary
  interactions at random times: the method of stochastic {Liouville} equation,
  and two examples of interactions in the context of a tight-binding chain.
\newblock {\em J. Stat. Mech.: Theory Exp.}, 2022(5):053101, 2022.

\bibitem{dattagupta_stochastic_2022}
S.~Dattagupta, D.~Das, and S.~Gupta.
\newblock Stochastic resets in the context of a tight-binding chain driven by
  an oscillating field.
\newblock {\em J. Stat. Mech.: Theory Exp.}, 2022(10):103210, 2022.

\bibitem{ishiyama2025exact_den}
T.~Ishiyama, K.~Fujimoto, and T.~Sasamoto.
\newblock Exact density profile in a tight-binding chain with dephasing noise.
\newblock {\em J. Stat. Mech.: Theor. Exp.}, 2025(3):033103, 2025.

\bibitem{ishiyama2025exact_curr}
T.~Ishiyama, K.~Fujimoto, and T.~Sasamoto.
\newblock Exact current fluctuations in a tight-binding chain with dephasing
  noise.
\newblock {\em arXiv preprint arXiv:2504.06989}, 2025.

\bibitem{touchette_large_2009}
H.~Touchette.
\newblock The large deviation approach to statistical mechanics.
\newblock {\em Phys. Rep.}, 478(1-3):1--69, 2009.

\bibitem{touchette_introduction_2018}
H.~Touchette.
\newblock Introduction to dynamical large deviations of {Markov} processes.
\newblock {\em Phys. A: Stat. Mech. Appl.}, 504:5--19, 2018.

\bibitem{jack_ergodicity_2020}
R.~L. Jack.
\newblock Ergodicity and large deviations in physical systems with stochastic
  dynamics.
\newblock {\em Eur. Phys. J. B}, 93(4):74, 2020.

\bibitem{zannetti_condensation_2014}
M.~Zannetti, F.~Corberi, and G.~Gonnella.
\newblock Condensation of fluctuations in and out of equilibrium.
\newblock {\em Phys. Rev. E}, 90(1):012143, 2014.

\bibitem{lepri_large-deviations_2024}
Stefano Lepri.
\newblock Large-deviations approach to thermalization: the case of harmonic
  chains with conservative noise.
\newblock {\em J. Stat. Mech.: Theory Exp.}, 2024(7):073208, 2024.

\bibitem{molina_nonlinear_1993}
M.~I. Molina and G.~P. Tsironis.
\newblock Nonlinear impurities in a linear chain.
\newblock {\em Phys. Rev. B}, 47(22):15330--15333, 1993.

\bibitem{tsironis_generalized_1994}
G.~P. Tsironis, M.~I. Molina, and D.~Hennig.
\newblock Generalized nonlinear impurity in a linear chain.
\newblock {\em Phys. Rev. E}, 50(3):2365--2368, 1994.

\bibitem{brazhnyi_spontaneous_2011}
V.~A. Brazhnyi and B.~A. Malomed.
\newblock Spontaneous symmetry breaking in {Schrödinger} lattices with two
  nonlinear sites.
\newblock {\em Phys. Rev. A}, 83(5):053844, 2011.

\bibitem{lepri_asymmetric_2011}
S.~Lepri and G.~Casati.
\newblock Asymmetric wave propagation in nonlinear systems.
\newblock {\em Phys. Rev. Lett.}, 106(16):164101, 2011.

\bibitem{dambroise_eigenstates_2013}
J.~D'Ambroise, P.~G. Kevrekidis, and S.~Lepri.
\newblock Eigenstates and instabilities of chains with embedded defects.
\newblock {\em Chaos}, 23(2):023109, 2013.

\bibitem{eilbeck_discrete_1985}
J.~C. Eilbeck, P.~S. Lomdahl, and A.~C. Scott.
\newblock The discrete self-trapping equation.
\newblock {\em Phys. D: Nonlinear Phenom}, 16(3):318--338, 1985.

\bibitem{rasmussen_statistical_2000}
K.~Ø. Rasmussen, T.~Cretegny, P.~G. Kevrekidis, and Niels Grønbech-Jensen.
\newblock Statistical mechanics of a discrete nonlinear system.
\newblock {\em Phys. Rev. Lett.}, 84(17):3740--3743, 2000.

\bibitem{kevrekidis_discrete_2009}
P.~G. Kevrekidis.
\newblock {\em The {Discrete} {Nonlinear} {Schrödinger} {Equation}:
  {Mathematical} {Analysis}, {Numerical} {Computations} and {Physical}
  {Perspectives}}.
\newblock Springer, Berlin, 2009.

\bibitem{flach_discrete_1998}
S.~Flach and C.~R. Willis.
\newblock Discrete breathers.
\newblock {\em Phys. Rep.}, 295(5):181--264, 1998.

\bibitem{flach_discrete_2008}
S.~Flach and A.~V. Gorbach.
\newblock Discrete breathers - {Advances} in theory and applications.
\newblock {\em Phys. Rep.}, 467(1-3):1--116, 2008.

\bibitem{johansson_dynamics_1998}
M.~Johansson, S.~Aubry, Y.~B. Gaididei, P.~L. Christiansen, and K.~Ø.
  Rasmussen.
\newblock Dynamics of breathers in discrete nonlinear schr{\"o}dinger models.
\newblock {\em Phys. D: Nonlinear Phenom.}, 119(1-2):115--124, 1998.

\bibitem{iubini_discrete_2013}
S~Iubini, R~Franzosi, R~Livi, G-L Oppo, and A~Politi.
\newblock Discrete breathers and negative-temperature states.
\newblock {\em New J. Phys.}, 15(2):023032, 2013.

\bibitem{iubini_effective_2025}
S.~Iubini and A.~Politi.
\newblock Effective grand canonical description of condensation in
  negative-temperature regimes.
\newblock {\em Phys. Rev. Lett.}, 134(9):097102, 2025.

\bibitem{politi_frozen_2022}
A.~Politi, P.~Politi, and S.~Iubini.
\newblock Frozen dynamics of a breather induced by an adiabatic invariant.
\newblock {\em J. Stat. Mech.: Theory Exp.}, 2022(4):043206, 2022.

\bibitem{iubini_dynamical_2019}
S.~Iubini, L.~Chirondojan, G.~Oppo, A.~Politi, and P.~Politi.
\newblock Dynamical freezing of relaxation to equilibrium.
\newblock {\em Phys. Rev. Lett.}, 122(8):084102, 2019.

\bibitem{rasmussen_localized_1998}
K.~Ø. Rasmussen, P.~L. Christiansen, M.~Johansson, Yu.~B. Gaididei, and S.~F.
  Mingaleev.
\newblock Localized excitations in discrete nonlinear schrödinger systems:
  Effects of nonlocal dispersive interactions and noise.
\newblock {\em Phys. D: Nonlinear Phenom.}, 113(2):134--151, 1998.

\bibitem{iubini_coarsening_2014}
S.~Iubini, A.~Politi, and P.~Politi.
\newblock Coarsening dynamics in a simplified {DNLS} model.
\newblock {\em J. Stat. Phys.}, 154(4):1057--1073, 2014.

\bibitem{iubini_relaxation_2017}
S.~Iubini, A.~Politi, and P.~Politi.
\newblock Relaxation and coarsening of weakly-interacting breathers in a
  simplified {DNLS} chain.
\newblock {\em J. Stat. Mech.: Theory Exp.}, 2017(7):073201, 2017.

\bibitem{ebrahimi_dynamics_2025}
M.~Ebrahimi, B.~Drossel, and W.~Just.
\newblock Dynamics of localised states in the stochastic discrete nonlinear
  {Schrödinger} equation.
\newblock {\em Phys. D: Nonlinear Phenom.}, 482:134905, 2025.

\bibitem{ebrahimi_stochastic_2025}
M.~Ebrahimi, B.~Drossel, and W.~Just.
\newblock The stochastic discrete nonlinear schr{\"o}dinger equation:
  microscopic derivation and finite-temperature phase transition.
\newblock {\em Phys. Scr.}, 101(13):135210, 2026.

\bibitem{acharya_dynamics_2026}
A.~Acharya, L.~Giuggioli, and S.~Gupta.
\newblock Dynamics and steady states of tight-binding chains in presence of
  isolated defects.
\newblock {\em J. Stat. Mech.: Theory Exp.}, 2026(4):043102, 2026.

\bibitem{perfect_spectral_1965}
H.~Perfect and L.~Mirsky.
\newblock Spectral properties of doubly-stochastic matrices.
\newblock {\em Monatsh. Math.}, 69(1):35--57, 1965.

\bibitem{schwarzer_moments_1972}
E.~Schwarzer and H.~Haken.
\newblock The moments of the coupled coherent and incoherent motion of
  excitons.
\newblock {\em Phys. Lett. A}, 42(4):317--318, 1972.

\bibitem{nazarenko_wave_2011}
S.~Nazarenko.
\newblock {\em Wave {Turbulence}}.
\newblock Springer, Berlin, 2011.

\bibitem{hannani_stochastic_2023}
A.~Hannani and S.~Olla.
\newblock A stochastic thermalization of the discrete nonlinear {Schrödinger}
  equation.
\newblock {\em Stoch. PDEs: Anal. Comp.}, 11(4):1379--1415, 2023.

\bibitem{lecomte_thermodynamic_2007}
V.~Lecomte, C.~Appert-Rolland, and F.~Van~Wijland.
\newblock Thermodynamic formalism for systems with markov dynamics.
\newblock {\em J. Stat. Phys.}, 127(1):51--106, 2007.

\bibitem{garrahan_dynamical_2007}
J.~P. Garrahan, R.~L. Jack, V.~Lecomte, E.~Pitard, K.~Van~Duijvendijk, and
  F.~Van~Wijland.
\newblock Dynamical first-order phase transition in kinetically constrained
  models of glasses.
\newblock {\em Phys. Rev. Lett.}, 98(19):195702, 2007.

\bibitem{garrahan_first-order_2009}
J.~P. Garrahan, R.~L. Jack, V.~Lecomte, E.~Pitard, K.~Van~Duijvendijk, and
  F.~Van~Wijland.
\newblock First-order dynamical phase transition in models of glasses: an
  approach based on ensembles of histories.
\newblock {\em J. Phys. A: Math. Theor.}, 42(7):075007, 2009.

\bibitem{corberi_probability_2019}
F.~Corberi and A.~Sarracino.
\newblock Probability distributions with singularities.
\newblock {\em Entropy}, 21(3):312, 2019.

\bibitem{gradenigo_first-order_2019}
G.~Gradenigo and S.~N. Majumdar.
\newblock A first-order dynamical transition in the displacement distribution
  of a driven run-and-tumble particle.
\newblock {\em J. Stat. Mech.: Theory Exp.}, 2019(5):053206, 2019.

\bibitem{maluckov_extreme_2009}
A.~Maluckov, Lj. Had\ifmmode~\check{z}\else \v{z}\fi{}ievski, N.~Lazarides, and
  G.~P. Tsironis.
\newblock Extreme events in discrete nonlinear lattices.
\newblock {\em Phys. Rev. E}, 79:025601(R), 2009.

\bibitem{maluckov_extreme_2013}
A.~Maluckov, N.~Lazarides, G.~P. Tsironis, and Lj. Had{\v{z}}ievski.
\newblock Extreme events in two-dimensional disordered nonlinear lattices.
\newblock {\em Physica D: Nonlinear Phenom.}, 252:59--64, 2013.

\bibitem{iubini_chain_2017}
S.~Iubini, S.~Lepri, R.~Livi, G.~Oppo, and A.~Politi.
\newblock A {Chain}, a {Bath}, a {Sink}, and a {Wall}.
\newblock {\em Entropy}, 19(9):445, 2017.

\bibitem{hoffmann_extreme_2017}
C.~{Hoffmann}, E.~G. {Charalampidis}, D.~J. {Frantzeskakis}, and P.~G.
  {Kevrekidis}.
\newblock Extreme events in near integrable lattices.
\newblock {\em arXiv e-prints. arXiv:1710.04899}, 2017.

\bibitem{kevrekidis_instabilities_2003}
P.~G. Kevrekidis, Y.~S. Kivshar, and A.~S. Kovalev.
\newblock Instabilities and bifurcations of nonlinear impurity modes.
\newblock {\em Phys. Rev. E}, 67(4):046604, 2003.

\bibitem{morales-molina_trapping_2006}
L.~Morales-Molina and R.~A. Vicencio.
\newblock Trapping of discrete solitons by defects in nonlinear waveguide
  arrays.
\newblock {\em Opt. Lett.}, 31(7):966, 2006.

\bibitem{palmero_solitons_2008}
F.~Palmero, R.~Carretero-González, J.~Cuevas, P.~G. Kevrekidis, and
  W.~Królikowski.
\newblock Solitons in one-dimensional nonlinear {Schrödinger} lattices with a
  local inhomogeneity.
\newblock {\em Phys. Rev. E}, 77(3):036614, 2008.

\bibitem{hennig_nonlinear_2025}
D.~Hennig.
\newblock Nonlinear discrete {Schrödinger} equations with a point defect.
\newblock {\em arXiv preprint arXiv:2412.10142}, 2025.

\bibitem{gradenigo_localization_2021}
G.~Gradenigo, S.~Iubini, R.~Livi, and S.~N. Majumdar.
\newblock Localization transition in the discrete nonlinear {Schrödinger}
  equation: ensembles inequivalence and negative temperatures.
\newblock {\em J. Stat. Mech.: Theory Exp.}, 2021(2):023201, 2021.

\bibitem{rumpf_simple_2004}
B.~Rumpf.
\newblock Simple statistical explanation for the localization of energy in
  nonlinear lattices with two conserved quantities.
\newblock {\em Phys. Rev. E}, 69(1):016618, 2004.

\bibitem{kenkre_self-trapping_1986}
V.~M. Kenkre and D.~K. Campbell.
\newblock Self-trapping on a dimer: {Time}-dependent solutions of a discrete
  nonlinear {Schrödinger} equation.
\newblock {\em Phys. Rev. B}, 34(7):4959--4961, 1986.

\bibitem{rumpf_intermittent_2004}
B.~Rumpf.
\newblock Intermittent movement of localized excitations of a nonlinear
  lattice.
\newblock {\em Phys. Rev. E}, 70(1):016609, 2004.

\bibitem{johansson_statistical_2004}
M.~Johansson and K.~Ø. Rasmussen.
\newblock Statistical mechanics of general discrete nonlinear {Schrödinger}
  models: {Localization} transition and its relevance for {Klein}-{Gordon}
  lattices.
\newblock {\em Phys. Rev. E}, 70(6):066610, 2004.

\bibitem{Note1}
The case $\alpha <1$ implies that the absolute defect energy increases slower
  than its norm $b$ for large $b$, a regime, which is beyond the scope of the
  present study.

\end{thebibliography}

\end{document}